\documentclass[useAMS,usenatbib]{mnras}    
\usepackage{amsmath} % Advanced maths commands

\usepackage{newtxtext,newtxmath}
\usepackage{hyperref}
\usepackage[T1]{fontenc}
\usepackage{ae,aecompl}
\usepackage{graphicx} % Including figure files
\usepackage[dvips]{epsfig}
\usepackage{float}
\usepackage[caption=false]{subfig}
\usepackage{subfloat}
\usepackage{textcomp}
\usepackage{tabularx}
\usepackage{wrapfig}
\usepackage{fixltx2e}
\usepackage{subfig}
\usepackage{siunitx}
\usepackage{afterpage}
\usepackage{appendix}
\usepackage{cleveref}

\graphicspath{{figures/}}
\title[Discovery of Giant Radio Galaxies from NVSS: Radio \& Infrared Properties]{Discovery of Giant Radio Galaxies from NVSS: Radio \& Infrared Properties}

\author[Pratik Dabhade et al.]{Pratik Dabhade$^{1,5}$,\thanks{E-mail: pratikdabhade13@gmail.com} Madhuri Gaikwad$^{2}$, Joydeep Bagchi$^{1}$, M. Pandey-Pommier$^{3}$, 
\newauthor Shishir Sankhyayan$^{4}$, Somak Raychaudhury$^{1}$
\\
$^{1}$Inter University Centre for Astronomy and Astrophysics (IUCAA), Pune University Campus, Pune 411007, India\\
$^{2}$National Centre for Radio Astrophysics, TIFR, Post Bag 3, Ganeshkhind, Pune - 411007, India\\
$^{3}$Univ Lyon, Univ Lyon1, Ens de Lyon, CNRS, Centre de Recherche Astrophysique de Lyon UMR5574, France\\
$^{4}$Indian Institute of Science Education and Research, Pune 411008, India\\
$^{5}$Leiden Observatory, Leiden University, Niels Bohrweg 2, 2333 CA, Leiden, Netherlands.}

\date{Accepted XXX. Received YYY; in original form ZZZ}

\pubyear{2017}

\begin{document}
\label{firstpage}
\pagerange{\pageref{firstpage}--\pageref{lastpage}}
\maketitle

\begin{abstract}

Giant radio galaxies (GRGs) are one of the largest astrophysical sources in the Universe with an overall projected linear size of $\sim$ 0.7 Mpc or more. Last six decades of radio astronomy research has led to the detection of thousands of radio galaxies. But only $\sim$ 300 of them can be classified as GRGs. The reasons behind their large size and rarity are unknown. We carried out a systematic search for these radio giants and found a  large  sample of  GRGs. In this paper, we report the discovery of 25 GRGs from NVSS, in the redshift range (z) $\sim$ 0.07 to 0.67. Their physical sizes range from $\sim$ 0.8 Mpc to $\sim$ 4 Mpc. Eight of these GRGs have sizes $\geq$ 2 Mpc which is a rarity. Here, for the first time, we investigate the mid-IR  properties of the optical hosts of the GRGs and classify them securely into various AGN types using the WISE mid-IR colours. Using radio and IR data, four of the hosts of GRGs  were observed to be radio loud quasars that extend up to 2 Mpc in radio size. These GRGs missed detection in earlier searches possibly because of their highly diffuse nature, low surface brightness and lack of optical data. The new GRGs are a significant addition to the existing sample that will contribute to better understanding of the physical properties of radio giants.

\end{abstract}


\vspace{-0.16in}
\section{Introduction}
Giant Radio Galaxies (hereafter GRGs) represent an extreme class of active galaxies which have their linear sizes from $\sim$ 0.7 to 5 Mpc thus placing them amongst the largest single astrophysical objects known to us. The first GRG 3C236 was discovered about 40 years ago by \citet{1974Natur.250..625W}, is $\sim$ 4.62 Mpc in size and still ranks among the largest known GRGs. Since then only a few other GRGs have been reported. The largest known GRG till date is J1420-0545 which spans  4.95 Mpc \citep{2008ApJ...679..149M}. Despite various studies on GRGs, no unified model has emerged that can explain the immense physical scale and other physical properties of these extreme class of radio sources. With the advent of large area sky surveys like NVSS (NRAO VLA Sky Survey) \citep{1998AJ....115.1693C}, WENSS (Westerbork Northern Sky Survey ) \citep{1997A&AS..124..259R} and SUMSS (Sydney University Molonglo Sky Survey) \citep{2003MNRAS.342.1117M} several  more GRGs have been found and many more await discovery through dedicated searches. Morphologically, most of the GRGs show bright hot spots at the end of their radio lobes and are hence classified as edge-brightened Fanaroff-Riley class II (FR-II) objects \citep{1974MNRAS.167P..31F}. These radio sources are born in the active  nuclei at the galactic centre. It is believed that their so called `central engine'  is a  mass accreting  super massive black hole (SMBH) of mass $10^{8} - 10^{10}$~$M_{\odot}$,  which is responsible for the ejection of the collimated, bipolar relativistic jets orthogonal to an accretion disc \citep{1969Natur.223..690L,1984RvMP...56..255B}. 
In common with normal radio galaxies, GRGs are usually hosted by bright elliptical galaxies. However, revealing an exceedingly  rare phenomenon, recently two  extremely massive and fast rotating spiral galaxies have been found that host relativistic jets and lobes that extend to megaparsec scales \citep{2011MNRAS.417L..36H,2014ApJ...788..174B}.  These extraordinary objects demonstrate that given the right conditions, the central engine of the spiral galaxies can launch megaparsec scale radio jets and form lobes around them.

Since the inception of radio astronomy  thousands of radio galaxies have been discovered, but only a tiny fraction of them ($\sim$ 300 so far) are GRGs
\citep{1996MNRAS.279..257S,1999MNRAS.309..100I,2001A&A...370..409L,2001A&A...374..861S,2001A&A...371..445M,2005AJ....130..896S,2006A&A...454...85M,2007AcA....57..227M,
2012MNRAS.426..851K}. Of these, only GRGs have been studied in sufficient detail in multiple wavebands for achieving a good understanding of 
their unusual nature. This puts a restriction on carrying out any statistical study of their properties. Understanding their birth \& evolution, the duty cycle of radio activity,
and the influence of the surrounding intergalactic medium (IGM), which confines the lobes far away from the host galaxy and provides a `working-surface' for the jets to act, are among the most important  problems in this field. Since the GRGs are known to extend to such large sizes, they are believed to be the last stop of radio galaxy evolution. Hence their study may help us impose important constraints on the various evolutionary models of radio galaxies.  For the existence of radio lobes and hot spots at large distances from the AGN (Active Galactic Nuclei), a low density, ambient IGM is implied in which the lobes expand and eventually diffuse out at the end of their lives. It was shown by \citet{2009MNRAS.393....2S} that GRGs can serve as outstanding probes of the IGM via the relationship between morphology of a radio galaxy and the properties of surrounding material which is part of the large-scale cosmic-web. GRGs can also transport enriched material from the host galaxy to large distances and pollute the IGM with non-thermal particles and magnetic fields \citep{1994RPPh...57..325K}. This non-thermal magnetized plasma may lurk in the IGM for billions of years, and become a promising source for injection of high energy ``seed'' particles in the turbulent intra-cluster medium (ICM) and in energetic shock waves associated with megaparsec scale central radio halos and peripheral radio relics, which are observed in some  merging galaxy clusters \citep{2001ASPC..250..454E,2006Sci...314..791B,2010Sci...330..347V}. Thus,  GRGs may also play an important role in the high  energy processes related to large-scale structure formation in the Universe. 

Despite many observations and  various theoretical models proposed, the mechanisms by which the relativistic jets are launched from the accretion discs around massive  black holes in AGNs arevery poorly understood \citep{1977MNRAS.179..433B,1982MNRAS.199..883B}. Till date, few theoretical models have been proposed to explain the growth of GRGs to such enormous sizes. One of the common explanation for their gigantic size is attributed to the less dense IGM around them, which allows them to expand and grow unhindered \citep{2008MNRAS.385.1286J,2008ApJ...677...63S,2009MNRAS.393....2S,2015MNRAS.449..955M}. This implies that GRGs are preferentially born in  sparser galactic environment. Another theory proposes that GRGs are very old radio galaxies \citep{1997MNRAS.292..723K} that have grown to their current sizes over a long timescales. Detailed multifrequency studies of larger sample of GRGs is required to  validate or challenge these models in addition to discovering more GRGs in diverse galactic environments.

It is also interesting to note that megaparsec scale radio emitting plasma lobes of GRGs are the largest known natural reservoirs of magnetic field and  non-thermal relativistic particles associated with a single galactic system. This makes GRGs excellent `calorimeters'  for measuring the energy output of the central black holes (BH),  because the giant radio lobes are believed to store most of the released energy of black holes  for a very long time interval and may also  reveal multiple  radio  episodes of black hole activity \citep{2001ApJ...560..178K}. Moreover, because the extended lobes of GRGs can contain charged particles long enough for them to be accelerated to extremely high
energies \citep{1984ARA&A..22..425H}, it has been suggested \citep{2004JKAS...37..343K,2009MNRAS.393.1041H} that shocks within the energetic jets and radio lobes of GRGs can be the possible sites for the acceleration of ultra high energy cosmic rays (UHECRs) whose origin is still unknown and a subject  of world-wide  research \citep{2008RvMA...20..198H,2009Natur.458..847B,2016arXiv160506504H}.\\

In this paper, we present a new  sample of GRGs, of linear sizes in the range of 0.8 Mpc to 3.7 Mpc which were discovered during our extensive search in the 1.4 GHz radio  images available in the NVSS. In the forthcoming papers, we will present more samples of GRGs along with results of statistical analysis and multi-wavelength studies of their properties. The upcoming sections are organised as follows: in section 2 we present  the optical, radio and near infrared analysis of  GRG sample along with the estimation of the mass of the  central black hole in  the  AGN. In section 3, we discuss  the individual GRGs and their physical properties.  In Section 4, we describe their collective properties and discuss important results and their wider astrophysical implications. Finally, in Section 5, we summarize our work and discuss the future possibilities. Throughout the paper we adopt the flat $\Lambda$CDM cosmological model based on the latest Planck results ($H_o$ = 67.8 km $s^{-1}$ 
$Mpc^{-1}$, $\Omega_m$ =0.308) \citep{2015arXiv150201589P}. We use it to determine the linear sizes, luminosities and other relevant physical parameters of the GRGs. The Ned Wright cosmology calculator \citep{2006PASP..118.1711W} was used to obtain parameters like comoving distance ($D_{c})$, luminosity distance ($D_{L})$ and 
scale (Kpc/$\arcsec$).
% Start Table 1 ------------------------------------------------------------------------------------------------------------------------
\begin{table*}
\begin{center}
\begin{minipage}{130mm}
\caption{Basic GRG information: z$^{\dagger}$ indicates spectroscopic redshifts.The linear projected size (Mpc) is computed using Eq.\ref{size1}. $S_{i(1.4GHz)}$ is integrated flux of the
source. The $P_{1.4GHz}$ is the Radio power computed using 
Eq.\ref{power}.}
\label{basicinfo}
\begin{tabular}{@{}cccccccclcll}
 \hline
  No. &    RA     &    DEC    & Angular Size &      z           & Linear Size & $S_{i(1.4GHz)}$ & $P_{1.4GHz}$ \\ 
     &   (J2000) &   (J2000) & ($\arcmin$)Arcmins   &                  &     (Mpc)   &     (mJy)       & $10^{25}W Hz^{-1}$ \\ 
\hline
GRG1 & 00:16:04.3 &$+$04:20:24.1 & 6.9 & $0.43281\pm0.00009^\dagger$ & 2.40 & $50\pm5$ & $1.7\pm0.2$ \\ 
GRG2 & 00:22:24.9 &$-$08:18:45.7 & 6.2 & $0.57147\pm0.00011^\dagger$ & 2.50 & $259\pm24$ & $14.6\pm1.3$ \\ 
GRG3 & 03:15:36.2 &$-$07:43:38.8 & 3.3 & $ 0.269\pm0.033$            & 0.86 & $151\pm16$ & $2.2\pm0.2$ \\ 
GRG4 & 04:29:25.8 &$+$00:33:04.8 & 6.6 & $0.468\pm0.092$             & 2.40 & $111\pm10$ & $4.5\pm0.7$ \\ 
GRG5 & 04:49:32.3 &$-$30:26:37.8 & 7.5 & $0.31496\pm0.00015^\dagger$  & 2.14 & $94\pm9$ & $1.8\pm0.2$ \\ 
GRG6 & 08:02:48.8 &$+$49:27:23.8 & 4.3 & $0.678\pm0.053$             & 1.88 & $52\pm7$ & $3.9\pm0.5$ \\ 
GRG7 & 08:57:01.7 &$+$01:31:30.9 & 6.2 & $0.27336\pm0.00007^\dagger$ & 1.58 & $99\pm9$ & $1.5\pm0.1$ \\ 
GRG8 & 10:58:38.6 &$+$24:45:35.1 & 10.4 & $0.201\pm0.016$              & 2.13 & $147\pm12$ & $1.3\pm0.1$ \\ 
GRG9 & 10:59:20.1 &$-$17:09:20.8 & 11.5 & $0.1027\pm0.0005^\dagger$   & 1.33 & $154\pm7$ & $0.4\pm0.01$ \\ 
GRG10 & 13:27:41.3 &$+$57:49:43.4 & 12.1 &$0.12018\pm0.00001^\dagger$  & 1.61 & $89\pm4$ & $0.3\pm0.009$ \\ 
GRG11 & 13:27:43.5 &$+$17:48:37.3 & 3.1 &$0.65688\pm0.00025^\dagger$  & 1.33 & $70\pm12$ & $4.9\pm0.8$ \\
GRG12 & 20:08:43.3 &$+$00:49:18.9 & 11 & $0.412\pm0.094$             & 3.71 & $81\pm3$ & $2.6\pm0.4$ \\ 
GRG13 & 20:34:49.2 &$-$26:30:36.4 & 6.7 & $0.10329\pm0.00015^\dagger$ & 0.78 & $76\pm6$ & $0.2\pm0.01$ \\ 
GRG14 & 20:59:39.8 &$+$24:34:23.9 & 8.2 & $0.116\pm0.019$            & 1.06 & $149\pm8$ & $0.4\pm0.02$ \\ 
GRG15 & 22:33:01.3 &$+$13:15:02.5 & 16 & $0.093\pm0.011$             & 1.71 & $125\pm4$ & $0.24\pm0.008$ \\ 
GRG16 & 22:50:39.1 &$+$28:44:45.5 & 8.4 & $0.097\pm0.009$            & 0.93 & $120\pm9$ & $0.2\pm0.01$ \\ 
GRG17 & 22:56:15.1 &$-$36:17:59.1 & 14.5 & $0.090252\pm0.00015^\dagger$& 1.51 & $195\pm11$ & $0.3\pm0.02$ \\ 
GRG18 & 23:04:44.8 &$-$10:50:48.2 & 4.3 & $0.21026\pm0.00018^\dagger$& 0.91 & $58\pm8$ & $0.5\pm0.07$ \\ 
GRG19 & 23:12:01.3 &$+$13:56:55.9 & 11.4 & $0.14041\pm0.00001^\dagger$& 1.74& $338\pm26$ & $1.4\pm0.1$ \\ 
GRG20 & 23:16:20.1 &$-$01:02:07.3 & 6.8 & $0.221\pm0.029$             & 1.50& $90\pm5$ & $0.9\pm0.07$ \\ 
GRG21 & 23:26:23.2 &$+$24:58:40.4 & 11.7 & $0.25497\pm0.00002^\dagger$& 2.88 & $274\pm27$ & $3.6\pm0.3$ \\ 
GRG22 & 23:28:49.9 &$-$08:25:12.7 & 4.7 & $0.555\pm0.056$             & 1.87 & $63\pm7$ & $3.4\pm0.3$ \\ 
GRG23 & 23:35:52.1 &$+$52:15:39.9 & 10.2 & $0.070649\pm0.00015^\dagger$ & 0.85 & $95\pm8$ & $0.10\pm0.008$ \\ 
GRG24 & 23:49:29.7 &$-$00:03:05.8 & 4.3 & $0.187\pm0.049$             & 0.84 & $28\pm3$ & $0.20\pm0.02$ \\ 
GRG25 & 23:55:31.6 &$+$02:56:07.1 & 8.1 & $0.657\pm0.235$             & 3.48 & $80\pm9$ & $5.7\pm1.7$ \\
\hline
\end{tabular}
\end{minipage}
\end{center}
\end{table*}
% End Table 1 ------------------------------------------------------------------------------------------------------------------------
\vspace{-0.19in}
\section {Multi-frequency Properties of GRG Sample}
In Table 1, we report the multi-frequency properties of our GRG sample. Henceforth, the GRGs are identified with
labels GRG1, GRG2, GRG3, etc.  Eight of the reported 25 GRGs have projected linear size $\geqslant$ 2 Mpc with the largest extending to $\sim 3.7$ Mpc which places them among the largest radio galaxies known so far. The basic information like sky position of optical host galaxy (RA and DEC), redshift, the angular (arcminutes) sizes, physical sizes (Mpc), radio flux density $S_{i(1.4GHz)}$ and power is  tabulated in Table~\ref{basicinfo}. Angular size of the GRGs was estimated using the available radio maps. The angular extent was taken to be the separation between the peaks (radio brightness) of the outermost lobe components and only sources with angular size $\geqslant$ 3$\arcmin$ were selected. The linear projected sizes  of the GRGs in Table~\ref{basicinfo} were  then computed using the  formula : 
\begin{equation}
 d(Mpc) = \frac{\theta \times D_c}{(1+z)} \times \frac{\pi}{10800}
 \label{size1}
\end{equation}
where \\
$\theta$ = Angle subtended by the source in sky in arcminutes.\\
z = Redshift.\\
$D_c$ = Comoving distance (Mpc).\\
d = Projected linear size of the source in Mpc.\\
\citep{1988gera.book.....K}.

\vspace{-0.19in}
\subsection{Identification of New GRGs and Radio Analysis}

All the 25 new GRGs were  discovered in the 1.4 GHz NVSS  images via extensive search for large scale double radio sources. This was done manually by careful visual inspection of NVSS radio maps. Search for more GRGs from NVSS is being carried out under our project \textit{\textbf{SAGAN}} \footnote{\url{https://sites.google.com/site/anantasakyatta/sagan}}. SAGAN stands for \textbf{S}earch $\&$ \textbf{A}nalysis of \textbf{G}RGs with \textbf{A}ssociated \textbf{N}uclei. In this project, for the first time, we aim to make a sample of all known GRGs and find new GRGs from the existing survey data. Our search has yielded more than 150 new giant radio galaxy candidates and we present only those confirmed as a GRG in this paper. The criteria to form the sample of 25 GRGs reported in this paper are:
\begin{enumerate}
 \item  Double radio sources with angular size $\geqslant$ 3 $\arcmin$.
 \item  Confirming the coincidence of radio core and host galaxy (as seen in optical) by overlaying radio and optical maps.
 \item Availability of reliable redshift (spectroscopic or photometric) information of the host galaxy from various optical surveys.
\end{enumerate}

Recently similar work with different methods was carried out by \citet{2016ApJS..224...18P} where the author identified radio sources $\geq 4\arcmin$ as giant radio source candidates via automated pattern recognition algorithms. Information such as host galaxy identification and redshift, which is imperative for confirming actual giant radio nature of a galaxy, is not supplemented in \citet{2016ApJS..224...18P}.

NVSS is a 1.4 GHz continuum survey covering the entire sky north of $-$40 deg declination \citep{1998AJ....115.1693C}. NVSS images are made with a relatively large restoring beam (45 $\arcsec$ FWHM) to yield the high surface-brightness sensitivity needed for completeness and photometric accuracy. Their rms brightness fluctuations are about 0.45 mJy/beam = 0.14 K (Stokes I) and 0.29 mJy/beam = 0.09 K (Stokes Q and U). The rms uncertainties in right ascension and declination vary from $<$ 1 $\arcsec$ for relatively strong (S $>$ 15 mJy) point sources to 7 $\arcsec$ for the faintest (S = 2.3 mJy) detectable sources. The completeness limit is about 2.5 mJy.

For a more  detailed  morphology of the  GRGs and  identification of AGN core, the VLA  high resolution FIRST (Faint Images of the Radio Sky at Twenty-cm) 
\citep{1995ApJ...450..559B} survey data was  employed. The FIRST survey at 1.4 GHz was done  using NRAO Very Large Array (VLA) in its B-configuration, using $2\times7$ 3-MHz frequency channels centered at 1365 and 1435 MHz. 
FIRST survey maps have 1.8 $\arcsec$ pixels, restoring beam of 5 $\arcsec$ and a typical rms of 0.15 mJy. FIRST data  for  GRG2, GRG3, GRG7, GRG8, GRG9, GRG20, GRG22, GRG24 and GRG25 is available.
In addition, we searched the SUMSS data and obtained the flux density of few GRGs at 843 MHz. SUMSS is an imaging survey \citep{2003MNRAS.342.1117M} of the sky south of declination, $\delta = - 30$ degrees at 843 MHz. The survey has a resolution of 45$\times$45 cosec$\mid\delta\mid$ arcsec$^{2}$. SUMSS is similar in sensitivity and  resolution to the northern  VLA NVSS. The SUMSS data is available only for GRG5 and GRG17, thereby providing their flux  density at 843 MHz.
The integrated flux densities ($S_{i}$) of  GRGs were computed using NVSS maps in Common Astronomy Software Applications (CASA) using the task \textquotedblleft CASA-VIEWER\textquotedblright. The radio colour maps overlaid with contours shown in Figure~\ref{nvss_C} for  GRGs were produced using the NVSS images.  For this purpose  \textit{Cubehelix} colour scheme was used to make  colour maps using task \textquotedblleft CASA-VIEWER\textquotedblright  in CASA.  The \textit{Cubehelix} colour scheme gives more weightage to green and yellow colours where human eye is more sensitive thereby bringing out essential details better  in the radio images \citep{2011BASI...39..289G}.

Radio powers  for GRGs were calculated using the formula: \citep{2009MNRAS.392..617D}
\begin{equation}
 P_{1.4} = 4\pi D_L^{2}S_{1.4} (1+z)^{\alpha - 1}
 \label{power}
\end{equation}
where $D_L$ is the luminosity distance, $S_{1.4}$ is the measured radio flux at 1.4 GHz,  $(1+z)^{\alpha - 1}$ is the standard k-correction  used in radio astronomy and $\alpha$ is the radio spectral index ($S_{\nu}$ $\propto$ $\nu^{-\alpha}$) for which we adopt a universal value of -1 which is usually observed  for radio galaxies. This value of $\alpha$ also takes care of the steepness for some sources.

\subsection{Optical Analysis}

The central regions of the GRGs were searched in the optical band for host galaxies and the brightest  galaxy  found  within  1 $\arcmin$ of the central peak was  identified as a host galaxy. In many cases the host galaxies were found to overlap with the compact radio cores  detected in NVSS and in some cases  in the VLA FIRST survey.  For optical images and spectroscopic redshifts we used  the Sloan Digital Sky Survey (SDSS) \citep{2000AJ....120.1579Y}, a major multi-filter imaging and spectroscopic
redshift survey using a dedicated 2.5-m wide-angle optical telescope at Apache Point Observatory. SDSS Data Release 12 (DR12) is the final data release of the SDSS-III, containing all SDSS observations till July 2014. The SDSS  also provides CCD  images in five filters ($\it u, g, r, i$ and $\it z$) and estimated photometric redshifts (photoZ), which were used  if  spectroscopic redshifts were unavailable.  
The photoZ redshifts are quoted for the host galaxies of  GRG3, GRG4, GRG6, GRG8, GRG12,  GRG14, GRG15, GRG16, GRG20, GRG22, GRG24 and GRG25. The SDSS  optical colour images of host galaxies are shown in  Figure~\ref{optical}. The spectroscopic redshifts of GRG1, GRG2, GRG7, GRG10, GRG11, GRG19 and GRG21 were obtained from SDSS DR12 \citep{2015ApJS..219...12A}. Host galaxies with spectroscopic information from SDSS also have velocity dispersion information which is described in detail in \citet{boltonvdisp}.

%This enabled us to compute the line flux ratios and excitation index according to \citet{2009A&A...495.1033B} and classify these GRGs into HERGs 
%(High excitation radio galaxies) and LERGs (Low excitation radio galaxies). Galaxies with excitation index (EI) greater than 0.95 are classified as HERGs.  
%\begin{equation}
%\begin{small}
%EI= Log(\frac{[OIII]}{H\beta})- \frac{1}{3}[ Log(\frac{[NII]}{H\alpha}) - Log(\frac{[SII]}{H\alpha}) + Log(\frac{[OI]}{H\alpha}) ] 
%\label{EI}
%\end{small}
%\end{equation}
%Unfortunately this criteria could only be adopted for GRG8 due to lack of appropriate line information.

For GRGs (GRG1, GRG2, GRG5, GRG7, GRG10 and GRG21) we adopted \citet{2012MNRAS.421.1569B} classification criteria to classify the hosts into HERGs (High Excitation Radio Galaxies) and LERGs (Low Excitation Radio Galaxies). If the host galaxy has [OIII] (5507 \AA) line equivalent width $>$ 5\AA,  it is classified as HERG else it is classified as LERG. This is the first time that such a study of hosts of GRGs has been attempted.

For some of the optical hosts (GRG5, GRG9, GRG13, GRG17 and GRG18)  the redshifts were obtained from the 6dF galaxy survey \citep{2009MNRAS.399..683J}. The 6dF refers to the Six Degree Field instrument survey  that uses optical fibers and robotic positioning technology on the Anglo-Australian 3.9m telescope.  Host spectrum of GRG23 was obtained from 2MASS (Two Micron All Sky Survey) redshift survey \citep{2012ApJS..199...26H}. For GRGs with spectroscopic host information not available in SDSS, computation of excitation index (EI) and corresponding classification was not attempted due to either incomplete information or lower SNR.

\subsection{Infrared Analysis}

In order to study and classify the AGN types for the GRGs we used the Wide-field Infrared Survey Explorer (WISE) \citep{2010AJ....140.1868W} mid-IR magnitudes and colours and AGN hosts of the GRGs. Colour-colour plots were used to complete this study and classification. The WISE survey is an all sky survey in four mid-IR bands [W1 (3.4$\mu$m), W2 (4.6$\mu$m), W3 (12$\mu$m), W4 (22$\mu$m)] with an angular resolution of 6.1, 6.4, 6.5 and 12 $\arcsec$ respectively. The WISE all-sky catalog was searched for all the host galaxies  of the GRGs in our sample within a search radius of 5 $\arcsec$ to obtain their magnitudes in the four mid-IR bands.
The mid-IR magnitudes allowed us to compute the luminosities in the four respective bands (Table~\ref{wise}).

\subsubsection{Mid Infrared: WISE COLOUR-COLOUR PLOT}

We used the mid-IR colours to obtain the properties of a possibly dust obscured galactic nucleus and determine the radiative efficiency of the AGN.  Mid-IR data is highly suitable for this purpose because the optical-UV radiation  from the AGN accretion disc is absorbed by the putative dusty torus and re-radiated in mid-IR bands. It has been shown that WISE colours can effectively distinguish  AGNs from star-forming and passive galaxies. Moreover, within the AGN subset itself, HERGs and LERGs stand out clearly on the mid-IR colour-colour and mid-IR-radio  plots \citep{2004ApJS..154..166L,2005ApJ...631..163S,2010ApJ...713..970A,2015MNRAS.449.3191Y}. For the first time in this paper, using  mid-IR  colours we  classify the GRGs into HERG, LERG, BLRG  (Broad Line Radio Galaxy) and QSO (quasar) categories.  \citet{2014MNRAS.438.1149G} have shown that W1-W2 versus W2-W4 plots display a better separation between LERGs and NLRGs (Narrow Line Radio 
Galaxies) than the W1-W2 versus W2-W3 plot. This in turn implies that the infrared emission  from the dusty  torus around an AGN is possibly  stronger in  W4 (22 $\mu$m)  band. We have produced the WISE mid-IR colour-colour plots (similar to \citep{2014MNRAS.438.1149G})  of radio loud AGN associated  with all our newly discovered GRGs.

\vspace{-0.05in}
\subsubsection{Near Infrared : Two Micron All Sky Survey (2MASS)}

The J (1.25 $\mu$m), H (1.65 $\mu$m) and K (2.17 $\mu$m) bands  near infrared magnitudes (Table~\ref{mass}) of all hosts of GRGs were obtained from the 2MASS catalog \citep{2006AJ....131.1163S}. The K band luminosities of the hosts of the GRGs were calculated using the K band magnitudes and redshifts. 
However, no data was available in 2MASS for GRG2, GRG6, GRG11, GR12, GRG15, GRG19, GRG20, GRG21, GRG22 and GRG25.

\subsection{Black Hole Mass Estimation}
The central black hole mass of the host of the GRGs was computed using two methods which are as follows:
\vspace{-0.15in}

\subsubsection{$M_{BH}$-$L_{K,bulge}$ relation}
The central black hole mass of GRGs as listed in Table~\ref{mass} was computed using the K band magnitude of the (bulge dominated)
elliptical host galaxy ($L_{K,bulge}$) and using  the  $M_{BH}$-$L_{K,bulge}$  correlation \citep{2003ApJ...589L..21M,2007MNRAS.379..711G}.
Absolute K band magnitude for the Sun of $M_{K,\sun}=3.28$ was used to compute the K band bulge luminosity.
\begin{equation}
\log(M_{\rm bh}/M_{\sun}) = 0.95(\pm0.15)\log \frac{L_{K,sph}}
 {10^{10.91} L_{K,\sun}}  + 8.26(\pm0.11) 
 \label{Eq_MD17K}
\end{equation}
where $L_{K,sph}/L_{K,\sun}$ is the K band luminosity of the spheroid 
component of the galaxy (i.e., the bulge or the elliptical galaxy itself)
in solar units. 
\vspace{-0.08in}

\subsubsection{$M_{BH}$-$\sigma$ relation}
Information on central  velocity dispersion for  optical hosts of GRG1, GRG2, GRG7, GRG10, GRG11, GRG19 and GRG21 is available via fiber spectroscopy in SDSS DR12 which (also listed in Table~\ref{msigma-mbh}) further  enabled us to compute the central black hole mass using the well known  $M_{BH}$-$\sigma$  relation \citep{2013ApJ...764..184M}
 \begin{equation}
 log (M_{BH} / M_{\odot}) = \alpha + \beta log (\sigma / 200 kms^{-1})
\end{equation}
where $\alpha = 8.32 \pm 0.05$ and $\beta = 5.64 \pm 0.32$.
%--------------------------------------------------------------------------------------------------------------------------------------

\vspace{-0.19in}
\section{Results}

In this section we provide notes on individual GRGs and discuss their optical, mid-IR and radio properties in detail.
\vspace{-0.05in}

\subsection{GRG1 (J001604+042024)}
GRG1 is a very low surface brightness radio source ($\sim$ 50 mJy)  with a projected linear size of about 2.4 Mpc. The radio maps from NVSS show two radio lobes on both sides of the faint core. The host galaxy is at redshift of 0.433 and is one of the distant GRGs in our sample. Based on the criteria suggested by \cite{2012MNRAS.421.1569B}, as mentioned in Section 2.2 we classify the host of GRG1 as LERG.

\subsection{GRG2 (J002224.96-081845)}
GRG2 is the fourth most distant GRG in our reporting sample with a spectroscopic redshift of 0.571. This results in an overall projected linear size of  $\sim$2.5 Mpc. 
A fewer fraction of GRGs with overall size greater than 2 Mpc are known and even fewer at higher redshifts. It is the most powerful source from our sample as shown in Table \ref{basicinfo}. The FIRST map clearly resolves the faint radio core and the two radio lobes. GRG2 can clearly be classified as a FR-II type radio galaxy. Based on the criteria suggested by \cite{2012MNRAS.421.1569B} as mentioned in Section 2.2 we classify the host of GRG2 as LERG. This GRG is also mentioned in \citet{2015salt.confE..34B} where they have obtained the spectroscopic redshift of the host galaxy which is similar to that of SDSS as quoted above.

\subsection{GRG3 (J031536-074338)}
GRG3 is one of the smallest GRGs in our sample with an overall projected linear size of about 0.86 Mpc. The FIRST map clearly resolves the radio core and lobes thereby confirming its FR-II type nature.

\subsection{GRG4 (J042925+003304): A Giant Radio Quasar?}
We observe GRG4 to have a very high infrared luminosity as seen in Table~\ref{wise}. GRG4 is located at a moderately high redshift of z $\sim$ 0.47 (SDSS-photoZ). 
Figure~\ref{optical} shows the host galaxy exhibiting a prominent red colour which could be attributed to obscuration from dust. This could be a possible reason why this galaxy has acquired such a large infrared luminosity.  Using the 2MASS K band bulge luminosity we computed the mass of the central black hole to be  massive at $\sim$ 3.5$\times 10^{9}$ $M_{\odot}$. Based on the high mass of the central black hole and high luminosities in radio and infrared bands, we infer GRG4 to be a quasar;  however high sensitivity  optical spectrum is still needed to confirm its quasar nature. Our claim is further supported by  its position on the WISE mid-IR colour-colour plots as seen in Figure~\ref{ccplot}.  The AGN clearly  lies in a region on the plot which is populated mostly by  radio loud quasars. The host galaxy also could be an ultra luminous infrared galaxy with its far infrared luminosity above the value of  $10^{12}$ $L_{\odot}$ as seen in Table~\ref{wise}. 

%------------------------------------------------------------------------------------------------------------------------
\begin{figure*}
\vbox{
\hbox{
\hspace{-0.9cm}
\subfloat[\textbf{GRG1 (Linear size = 2.40 Mpc)}\label{subfig-1}]{%
\includegraphics[height=4.2in,width=4in]{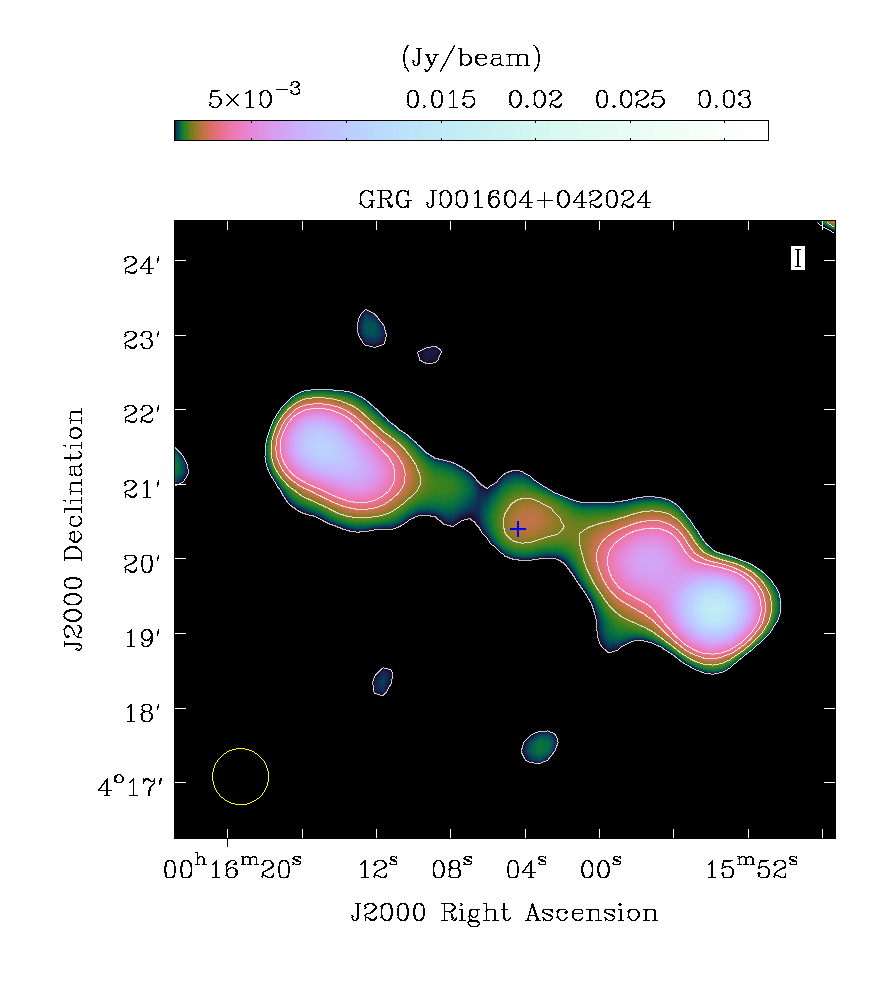}
}
\hspace{-0.2cm}
\subfloat[\textbf{GRG2 (Linear size = 2.50 Mpc)}\label{subfig-2}]{%
\includegraphics[height=4.2in,width=4in]{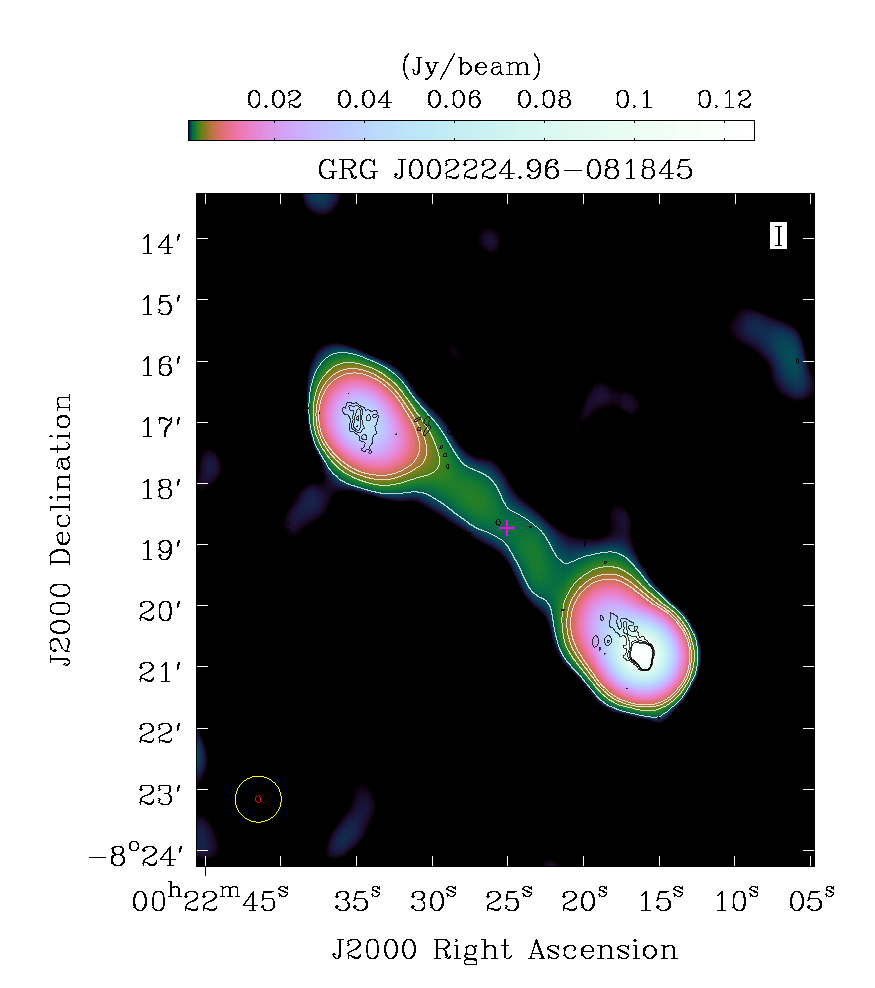}
}
\vspace{0.4in}
}
\hbox{
\hspace{-0.9cm}
\subfloat[\textbf{GRG3 (Linear size = 0.86 Mpc)}\label{subfig-3}]{%
\includegraphics[height=4.2in,width=4in]{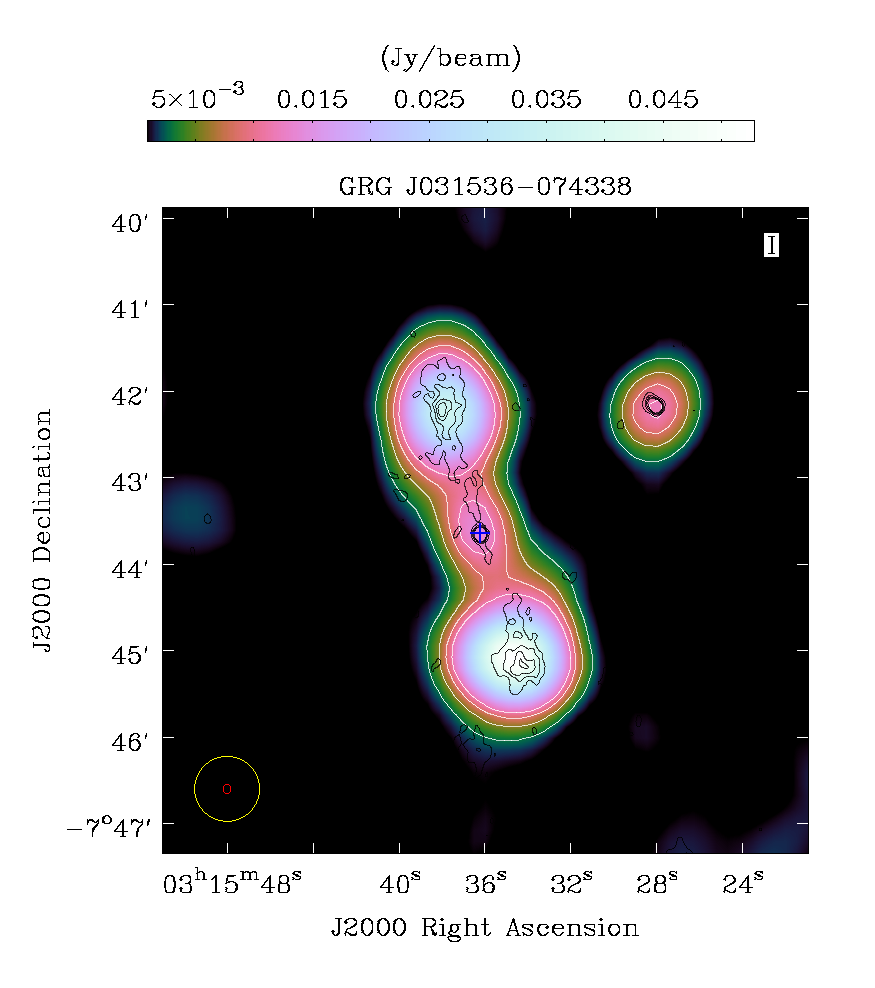}
}
\hspace{-0.2cm}
\subfloat[\textbf{GRG4 (Linear size = 2.40 Mpc)}\label{subfig-4}]{%
\includegraphics[height=4.2in,width=4in]{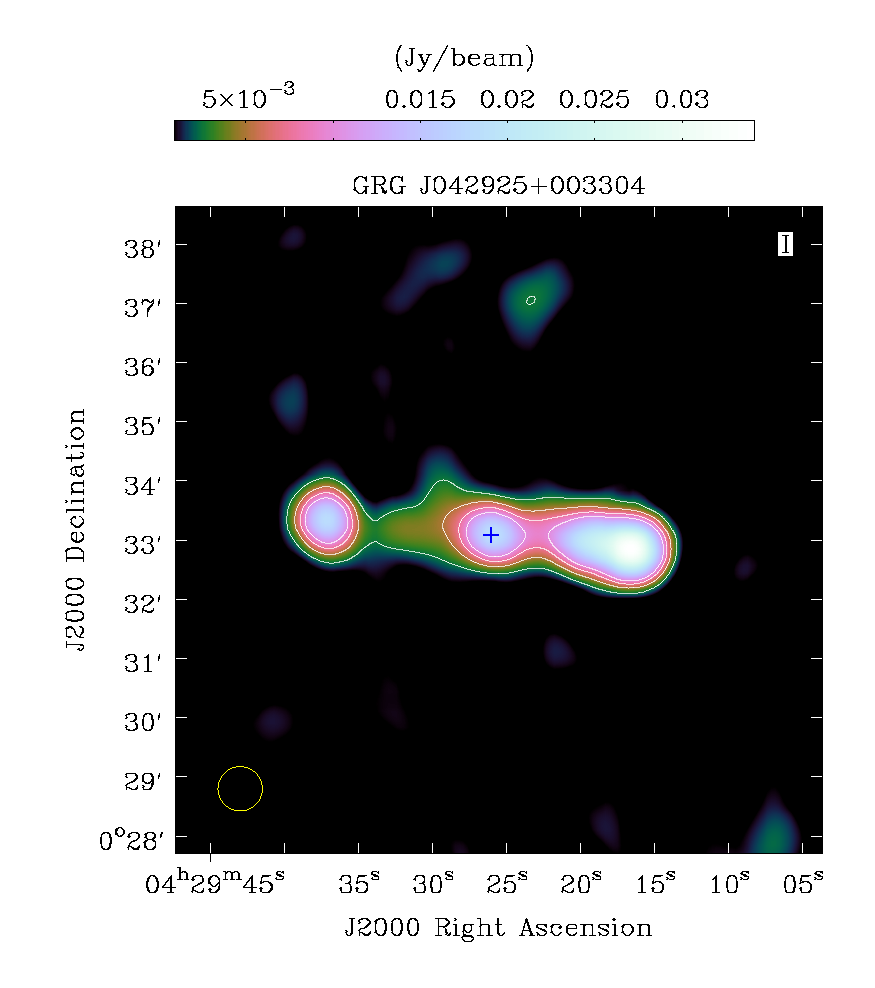}
}
\vspace{0.5in}
}
}
\caption{}
\end{figure*}

%------------------------------------------------------------------------------------------------------------------
\begin{figure*}
\ContinuedFloat
\vbox{
\hbox{
\hspace{-0.9cm}
\subfloat[\textbf{GRG5 (Linear size = 2.14 Mpc)}\label{subfig-5}]{%
\includegraphics[height=4.2in,width=4in]{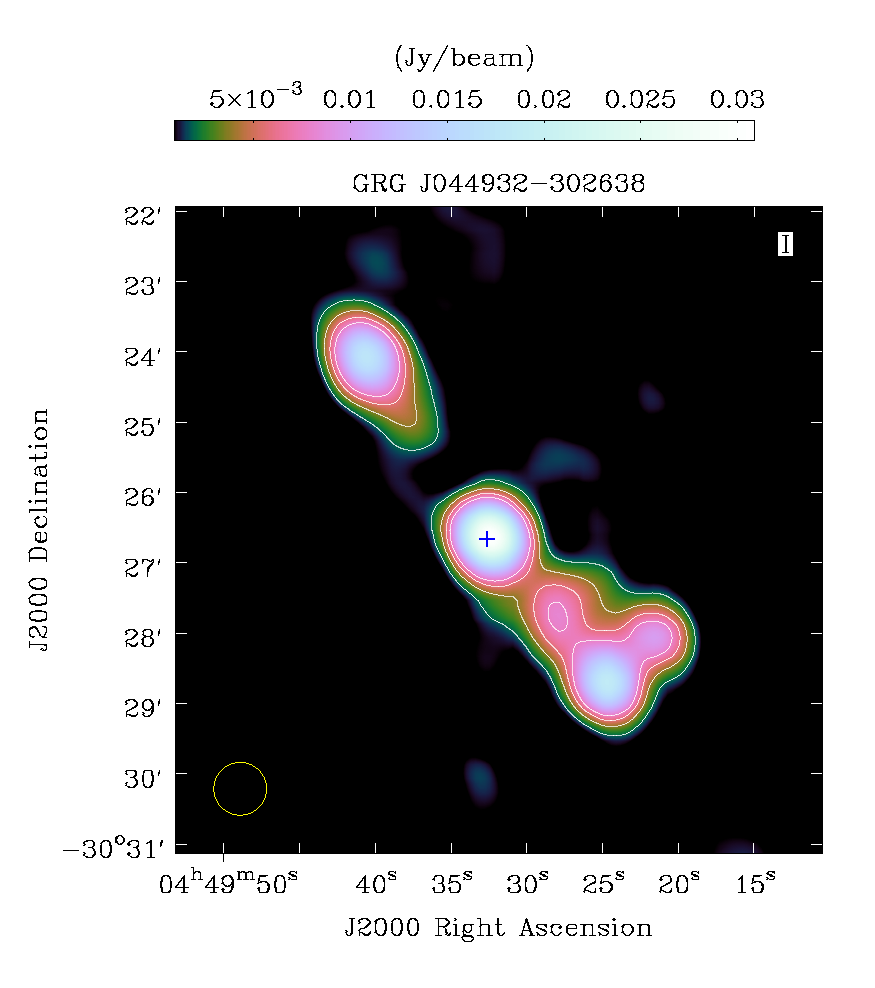}
}
\hspace{-0.5cm}
\subfloat[\textbf{GRG6 (Linear size = 1.88 Mpc)}\label{subfig-6}]{%
\includegraphics[height=4.2in,width=4.1in]{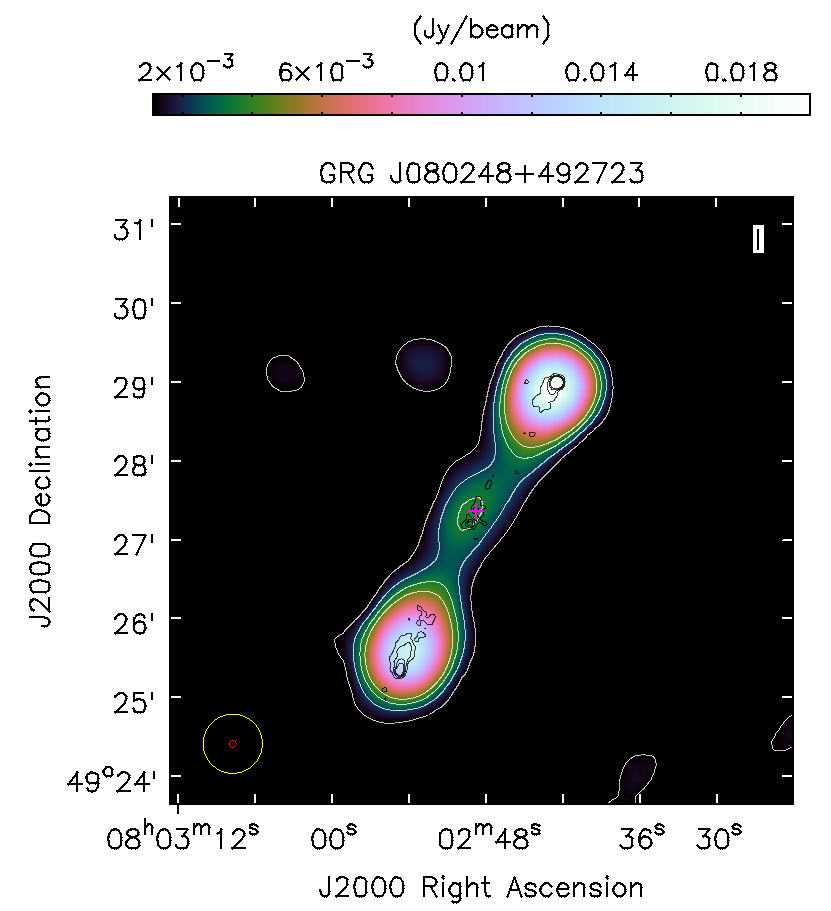}
}
\vspace{0.4in}
}
\hbox{
\hspace{-0.9cm}
\subfloat[\textbf{GRG7 (Linear size = 1.58 Mpc)}\label{subfig-7}]{%
\includegraphics[height=4.2in,width=4in]{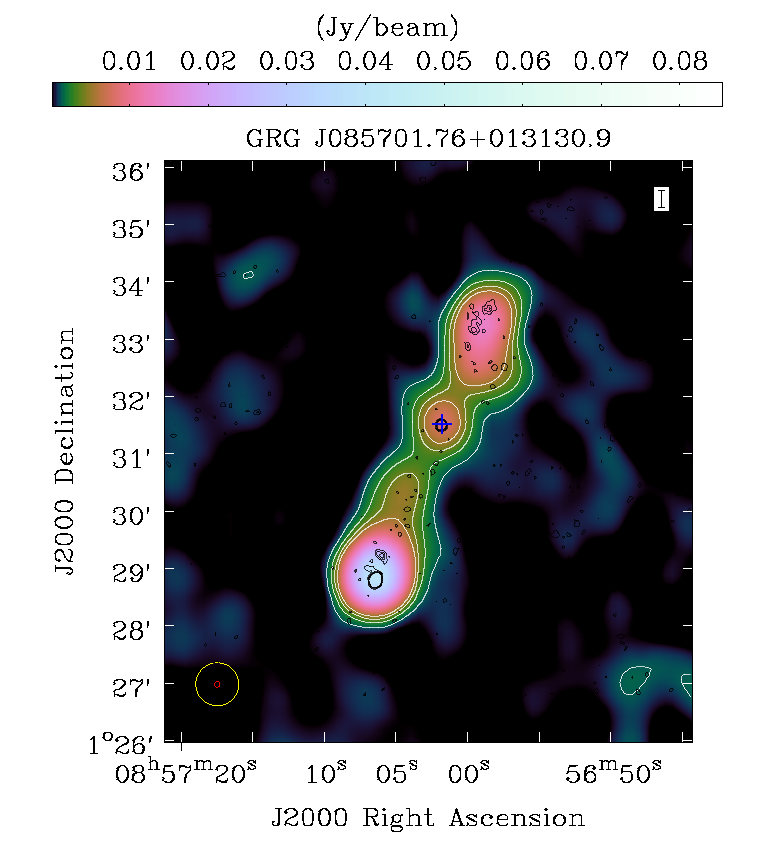}
}
\hspace{-0.36cm}
\subfloat[\textbf{GRG8 (Linear size = 2.13 Mpc)}\label{subfig-8}]{%
\includegraphics[height=4.36in,width=4.25in]{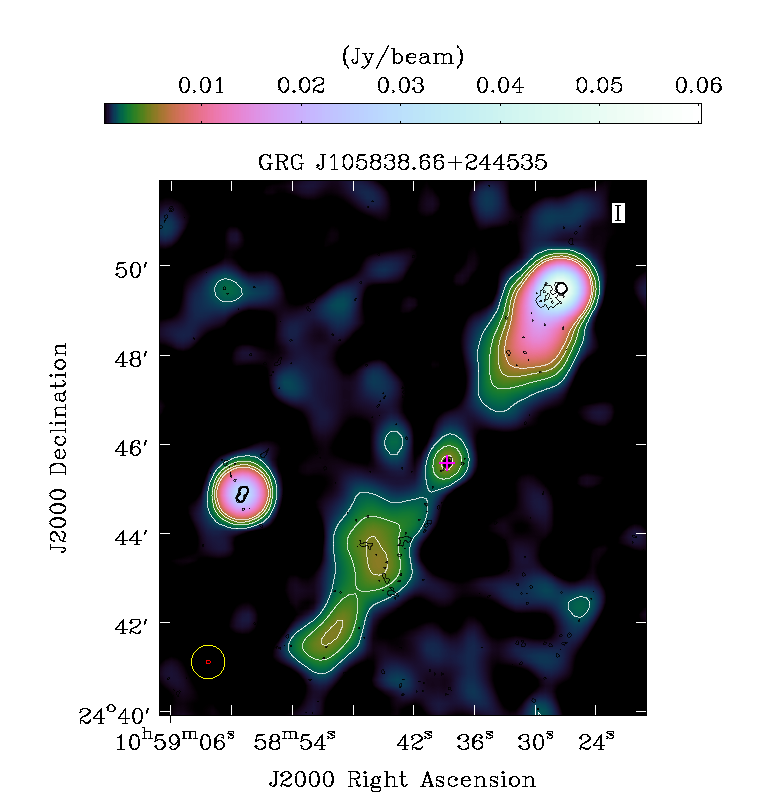}
}
\vspace{0.5in}
}
}
\caption{}
\end{figure*}
%------------------------------------------------------------------------------------------------------------------

\begin{figure*}
\ContinuedFloat
 \vbox{
\hbox{
\hspace{-0.9cm}
\subfloat[\textbf{GRG9 (Linear size = 1.33 Mpc)}\label{subfig-9}]{%
\includegraphics[height=4.2in,width=4in]{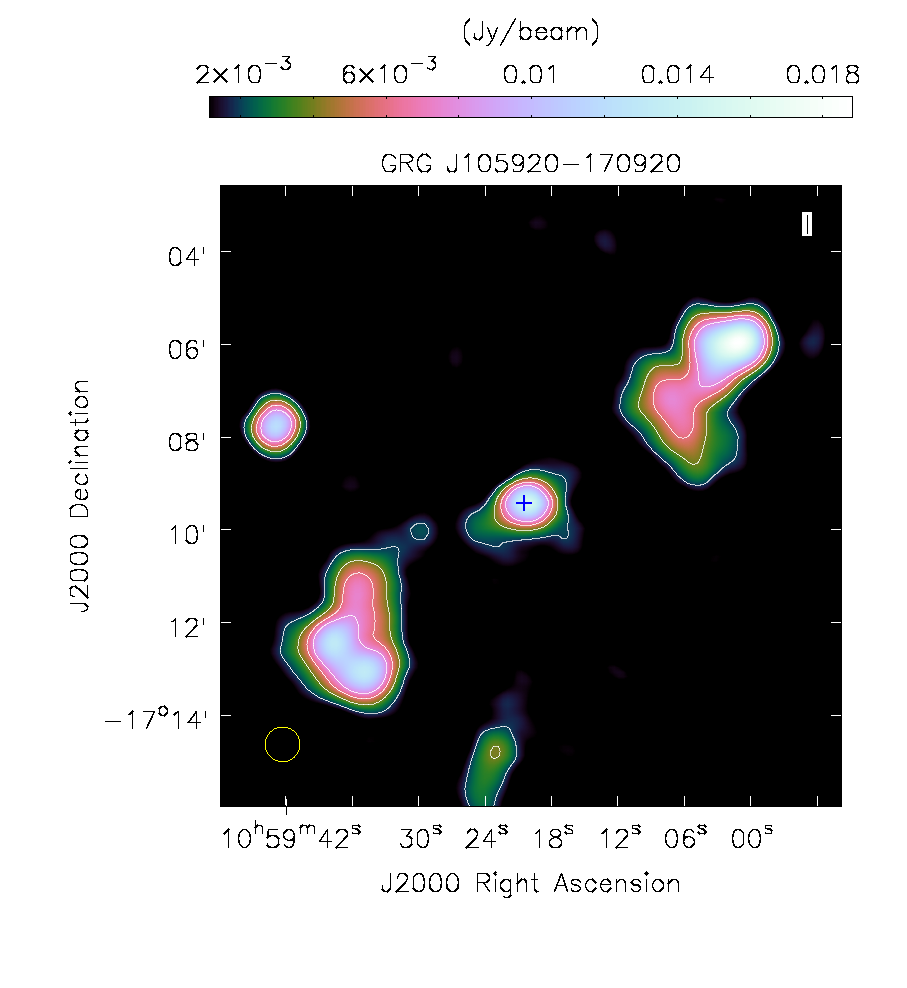}
}
\hspace{-0.2cm}
\subfloat[\textbf{GRG10 (Linear size = 1.61 Mpc)}\label{subfig-10}]{%
\includegraphics[height=4.2in,width=4in]{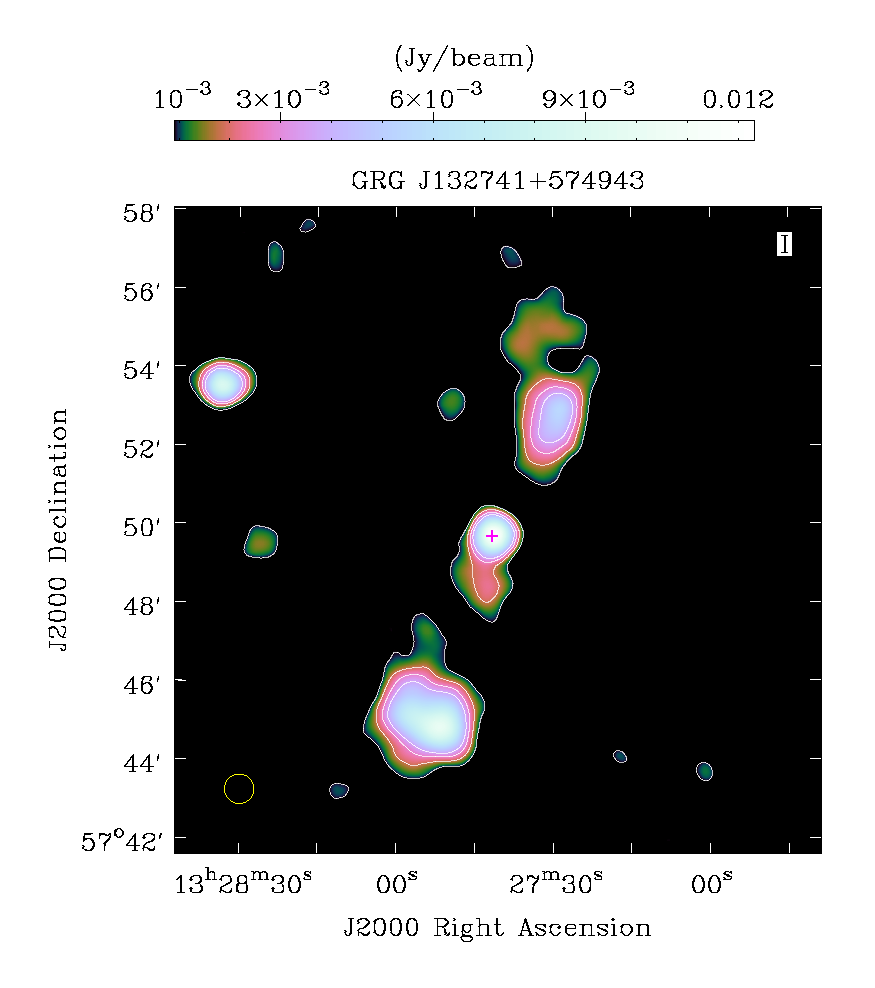}
}
\vspace{0.4in}
}
\hbox{
\hspace{-0.9cm}
\subfloat[\textbf{GRG11 (Linear size = 1.33 Mpc)}\label{subfig-11}]{%
\includegraphics[height=4.2in,width=4in]{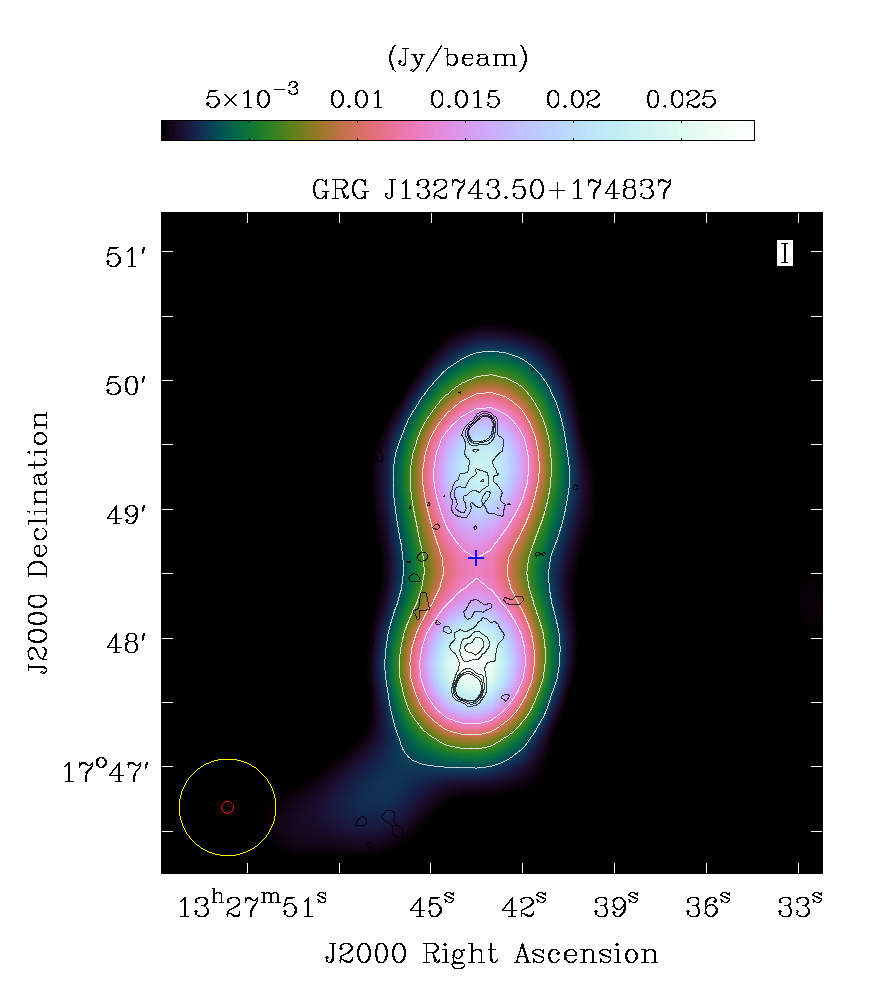}
}
\hspace{-0.2cm}
\subfloat[\textbf{GRG12 (Linear size = 3.71 Mpc)}\label{subfig-12}]{%
\includegraphics[height=4.2in,width=4in]{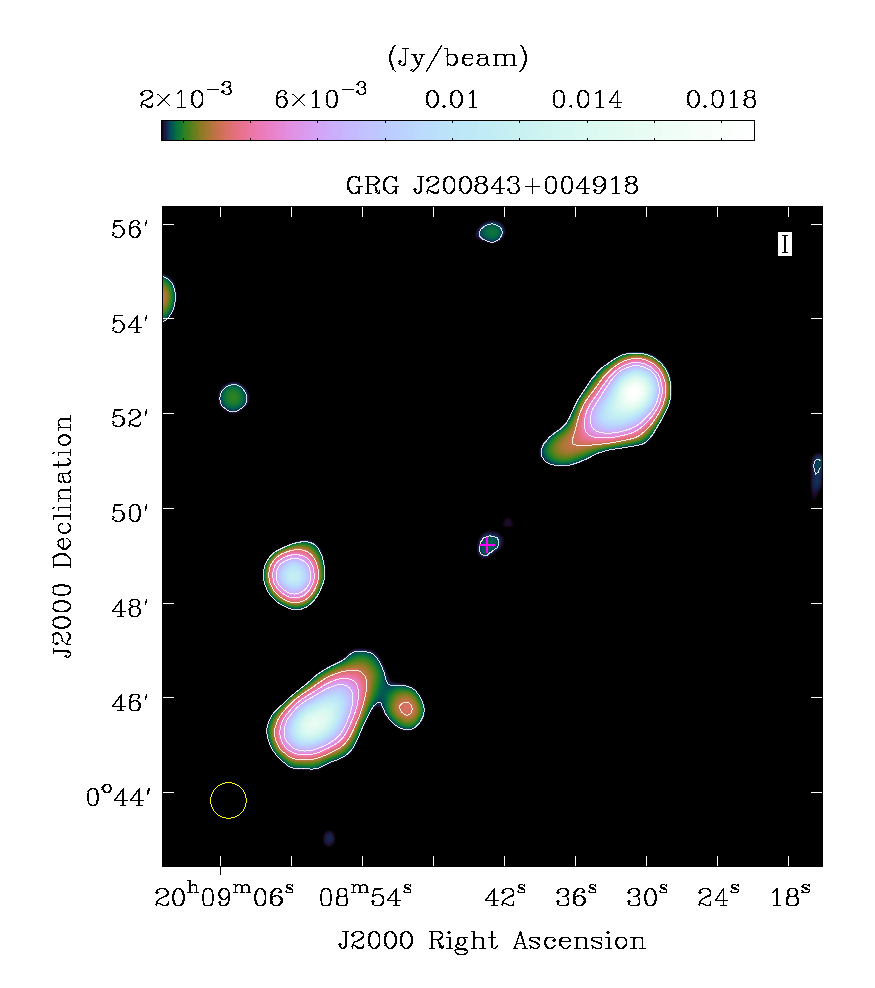}
}
\vspace{0.5in}
}
}
\caption{}
\end{figure*}
%------------------------------------------------------------------------------------------------------------------

\begin{figure*}
\ContinuedFloat
\vbox{
\hbox{
\hspace{-0.9cm}
\subfloat[\textbf{GRG13 (Linear size = 0.78 Mpc)}\label{subfig-13}]{%
\includegraphics[height=4.2in,width=4in]{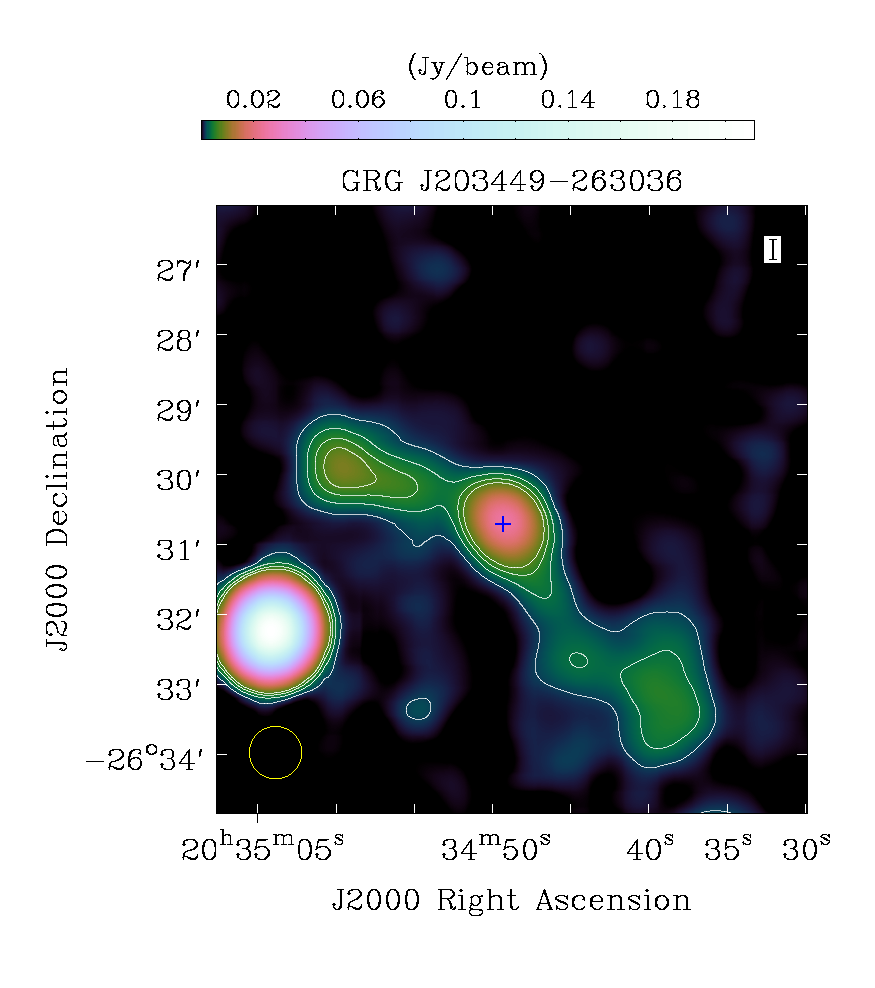}
}
\hspace{-0.2cm}
\subfloat[\textbf{GRG14 (Linear size = 1.06 Mpc)}\label{subfig-14}]{%
\includegraphics[height=4.2in,width=4in]{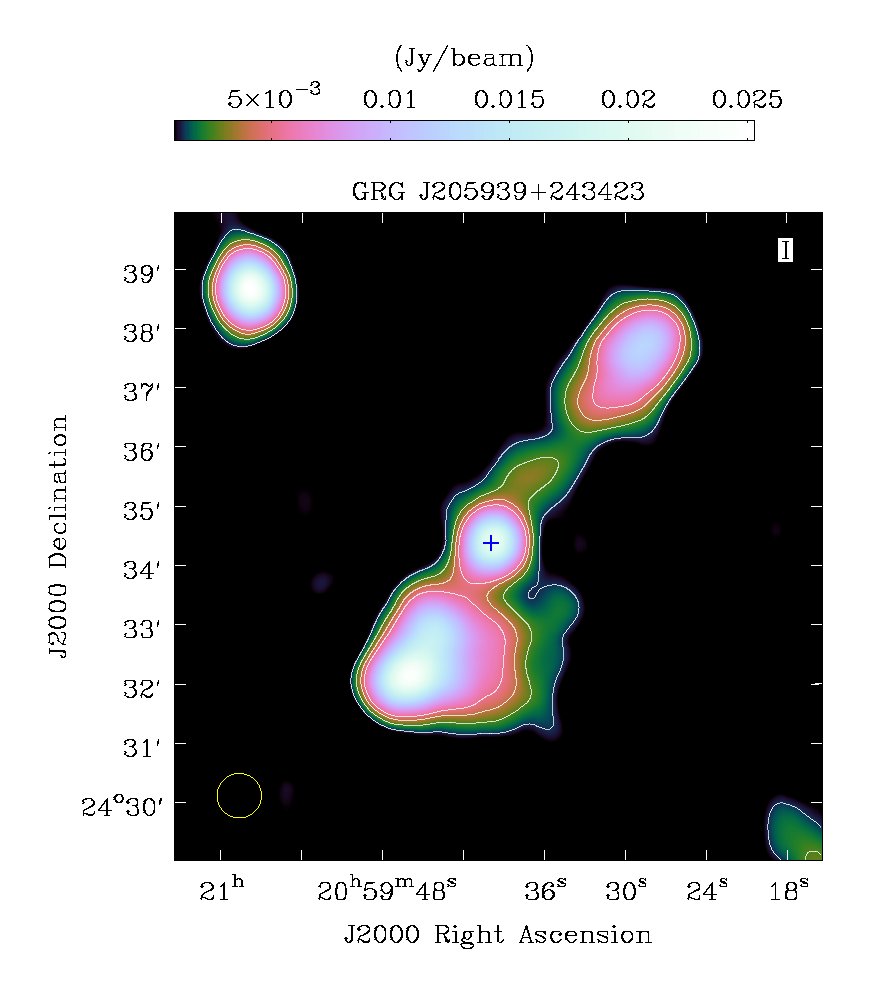}
}
\vspace{0.4in}
}
\hbox{
\hspace{-0.9cm}
\subfloat[\textbf{GRG15 (Linear size = 1.71 Mpc)}\label{subfig-15}]{%
\includegraphics[height=4.2in,width=4in]{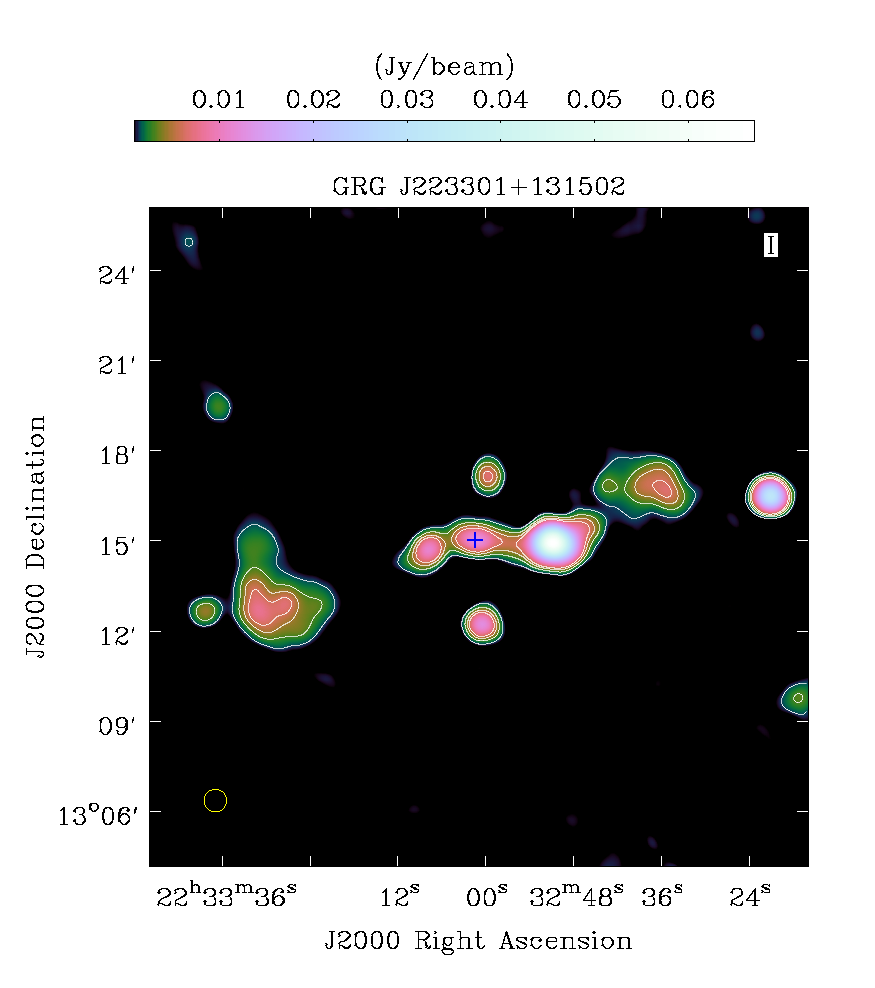}
}
\hspace{-0.2cm}
\subfloat[\textbf{GRG16 (Linear size = 0.93 Mpc)}\label{subfig-16}]{%
\includegraphics[height=4.2in,width=3.85in]{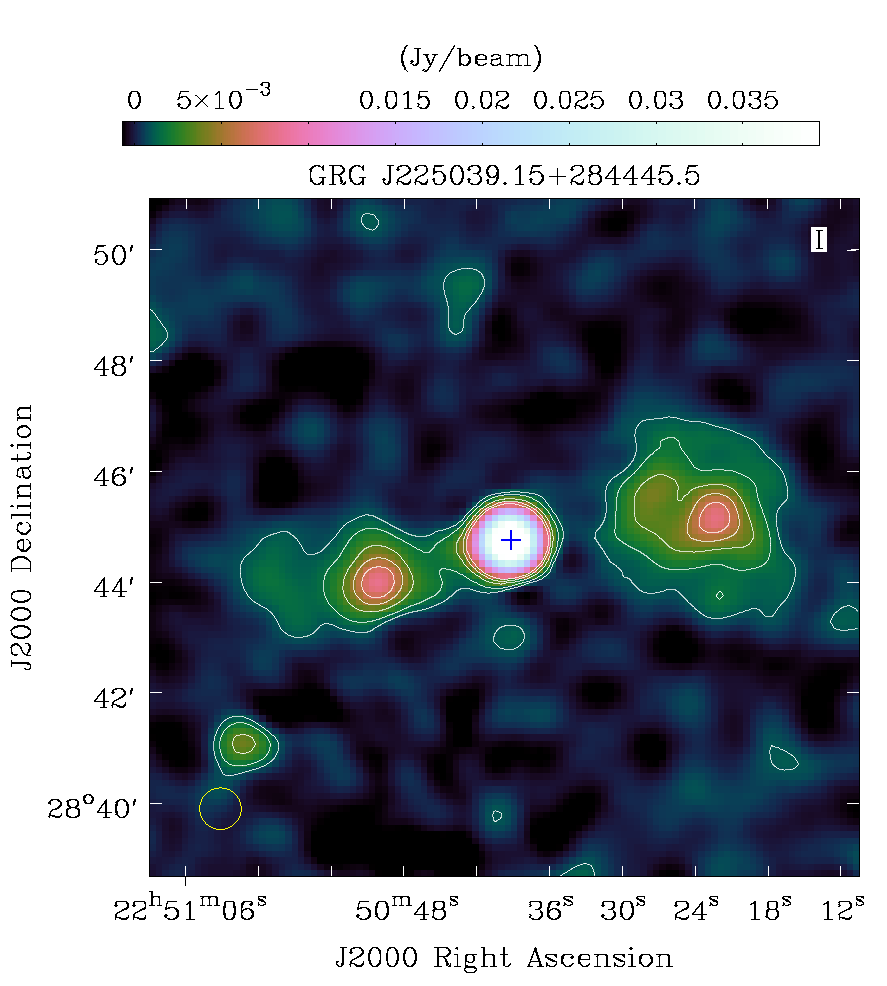}
}
}
\vspace{0.5in}
}
\caption{}
\end{figure*}
%------------------------------------------------------------------------------------------------------------------

\begin{figure*}
\ContinuedFloat
\vbox{
\hbox{
\hspace{-0.9cm}
\subfloat[\textbf{GRG17 (Linear size = 1.51 Mpc)}\label{subfig-17}]{%
\includegraphics[height=4.2in,width=4in]{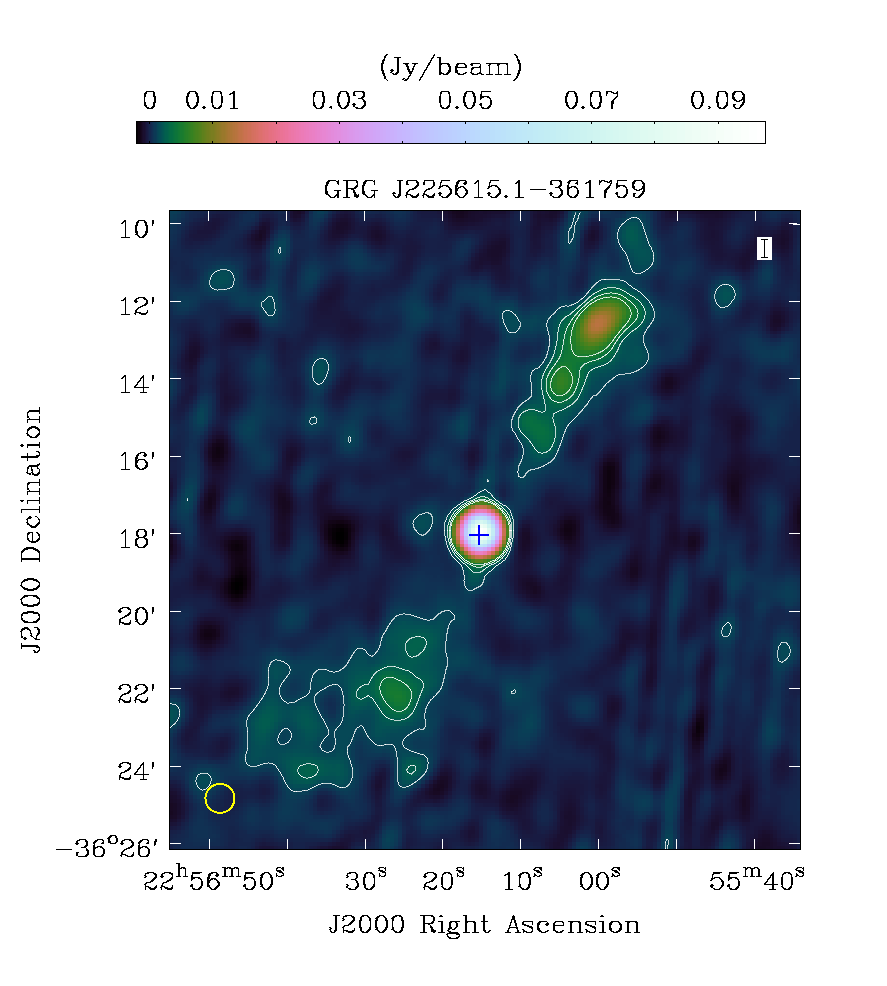}
}
\hspace{-0.2cm}
\subfloat[\textbf{GRG18 (Linear size = 0.91 Mpc)}\label{subfig-18}]{%
\includegraphics[height=4.2in,width=4in]{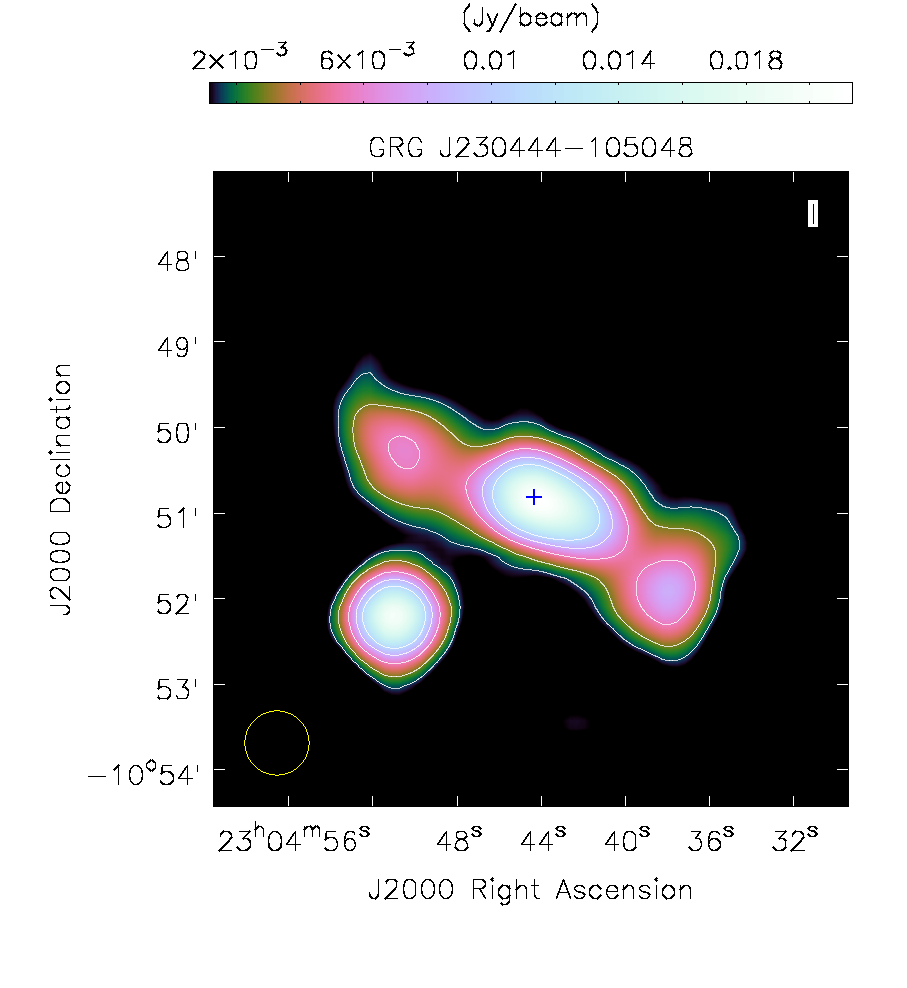}
}
\vspace{0.4in}
}
\hbox{
\hspace{-1.0cm}
\subfloat[\textbf{GRG19 (Linear size = 1.74 Mpc)}\label{subfig-19}]{%
\includegraphics[height=4.2in,width=4.2in]{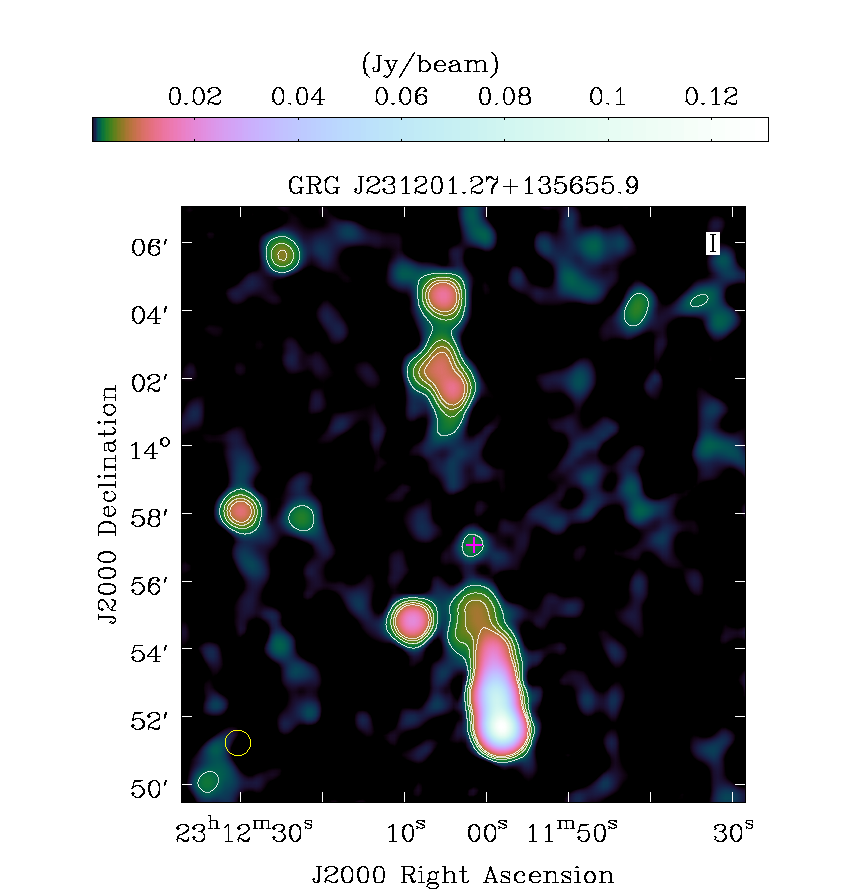}
}
\hspace{-0.39cm}
\subfloat[\textbf{GRG20 (Linear size = 1.50 Mpc)}\label{subfig-20}]{%
\includegraphics[height=4.1in,width=3.9in]{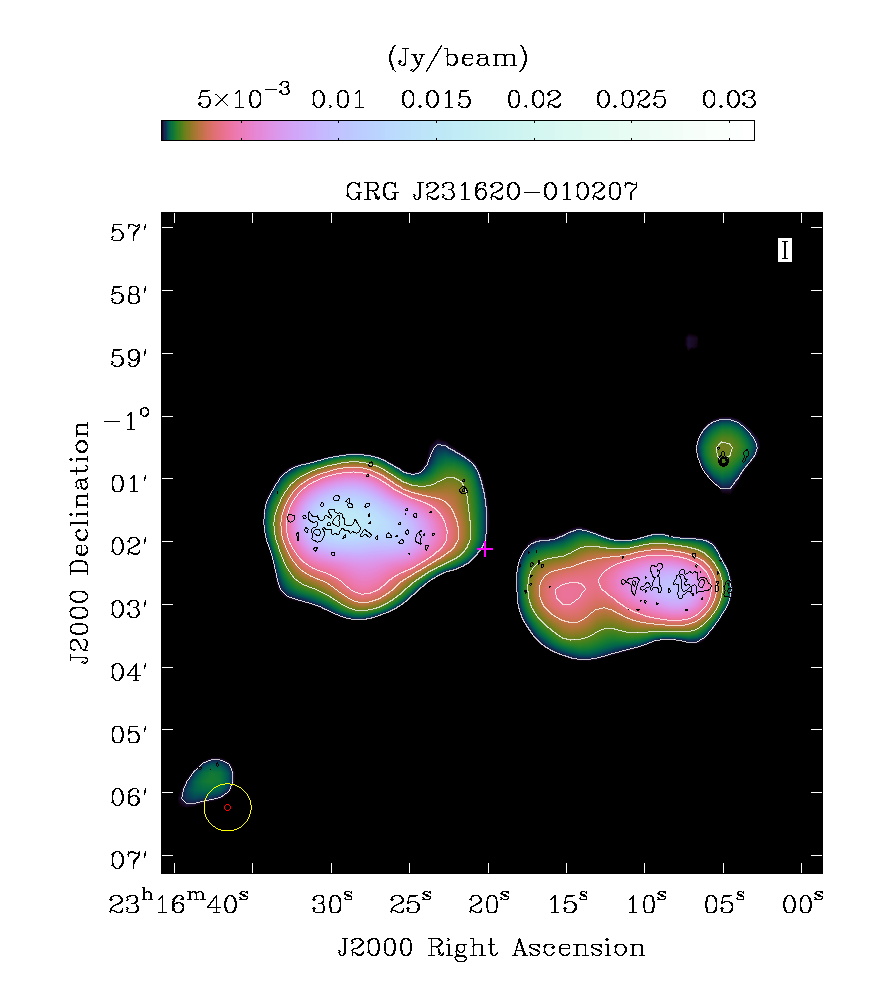}
}
}
\vspace{0.5in}
}
\caption{}
\end{figure*}

%------------------------------------------------------------------------------------------------------------------
\begin{figure*}
\ContinuedFloat
\vbox{
\hbox{
\hspace{-0.65cm}
\vspace{-1cm}
\subfloat[\textbf{GRG21 (Linear size = 2.88 Mpc)}\label{subfig-21}]{%
\includegraphics[height=4.21in,width=4in]{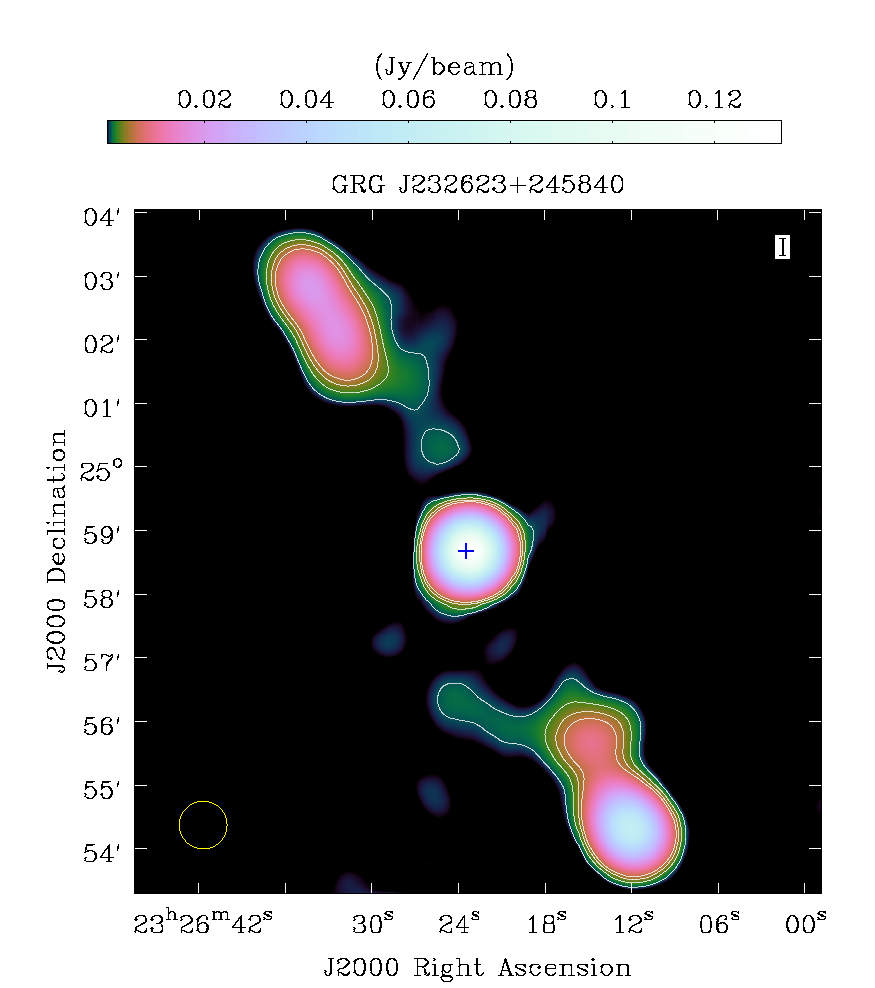}
}
\hspace{-0.3cm}
\subfloat[\textbf{GRG22 (Linear size = 1.87 Mpc)}\label{subfig-22}]{%
\includegraphics[height=4.2in,width=3.9in]{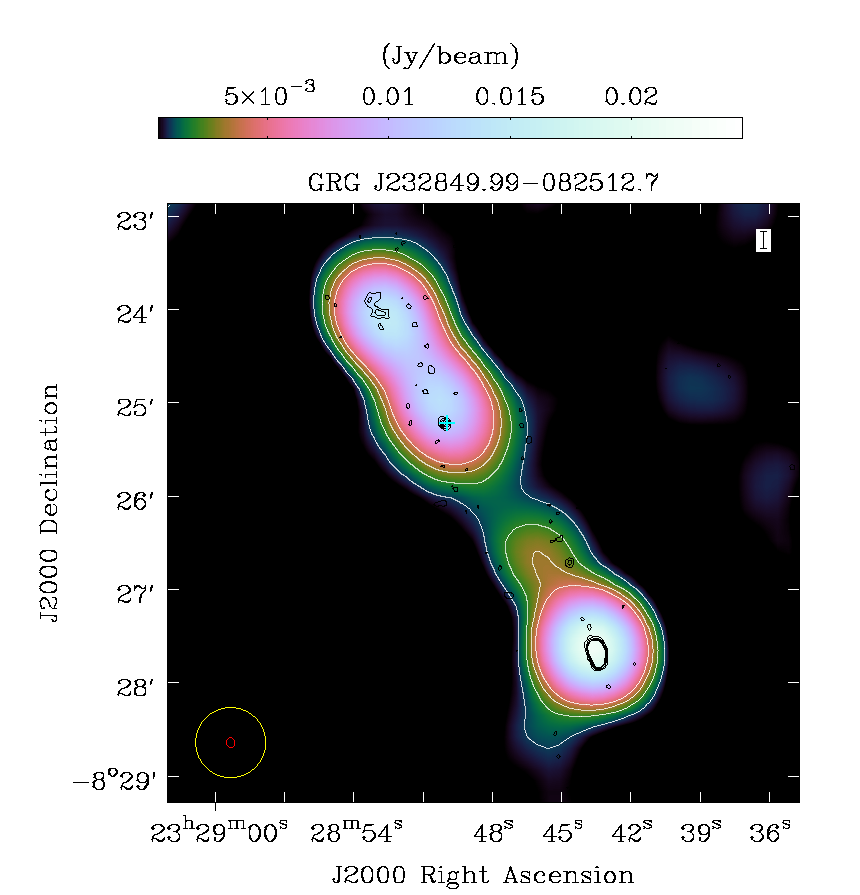}
}
\vspace{0.4in}
}
\hbox{
\hspace{-0.9cm}
\subfloat[\textbf{GRG23 (Linear size = 0.85 Mpc)}\label{subfig-23}]{%
\includegraphics[height=4.2in,width=4in]{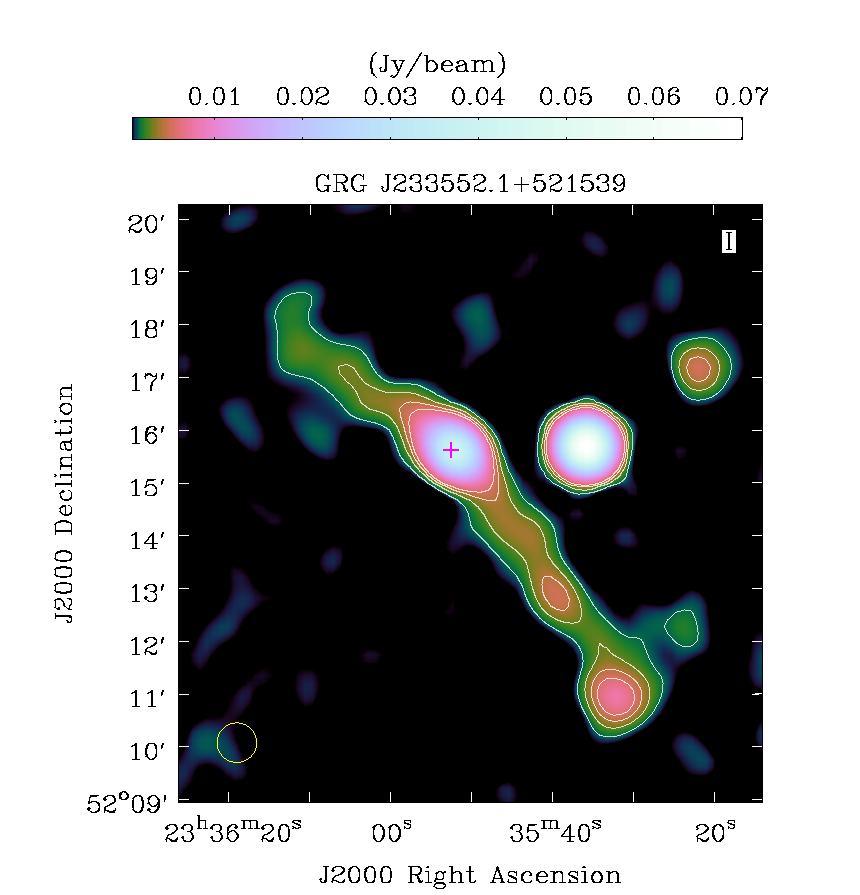}
}
\hspace{-0.2cm}
\subfloat[\textbf{GRG24 (Linear size = 0.84 Mpc)}\label{subfig-24}]{%
\includegraphics[height=4.2in,width=4in]{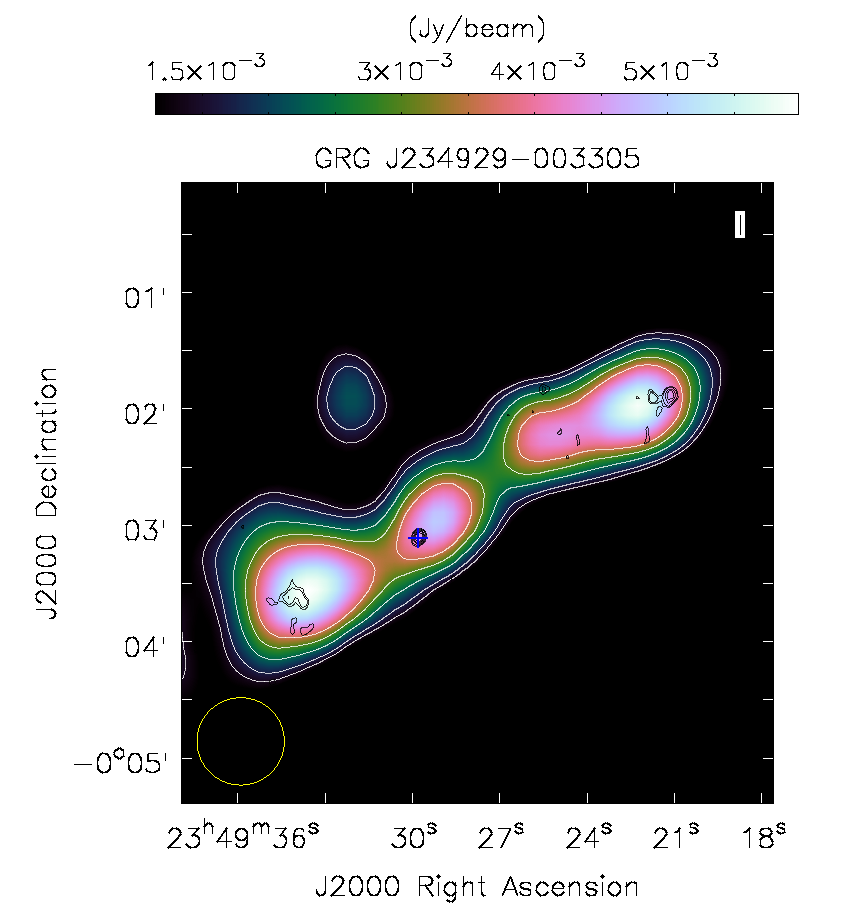}
}
}
\vspace{0.2in}
}
\caption{}
\end{figure*}
%------------------------------------------------------------------------------------------------------------------

\begin{figure*}
\ContinuedFloat
\vbox{
\hspace{2.4cm}
\subfloat[\textbf{GRG25 (Linear size = 3.48 Mpc)}\label{subfig-25}]{%
\includegraphics[height=4.2in,width=4in]{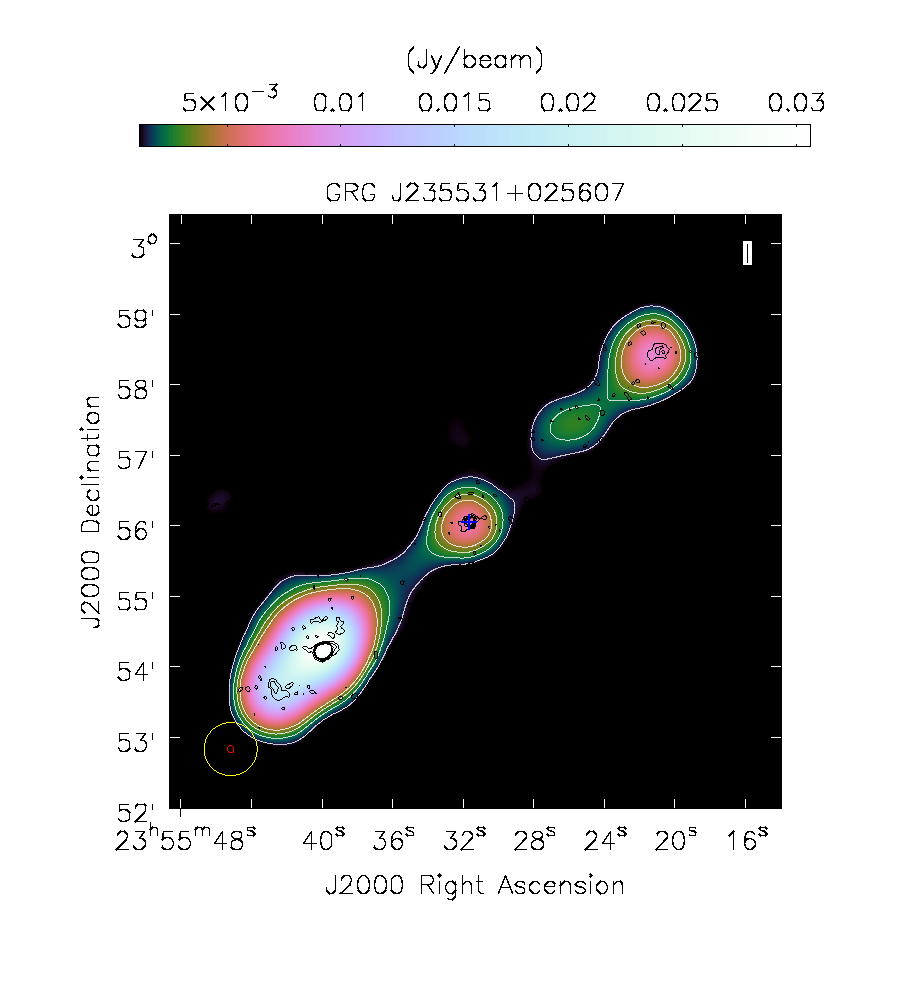}
}
\vspace{0.2in}
}
\caption{NVSS colour (Cube-Helix) \& contour maps of GRG1 - GRG25 made using task \textquotedblleft viewer\textquotedblright~in CASA with relative contour levels as [0.2, 0.4, 0.6, 0.8] mJy/beam. The beam size (45$\arcsec\times45\arcsec$ FWHM) for NVSS is represented in yellow colour and in red for FIRST (5$\arcsec\times5\arcsec$ FWHM) at the bottom left corner of each image. The white contours are of NVSS and the black contours (relative contour levels as [0.2, 0.4, 0.6, 0.8] mJy/beam) are of FIRST survey. The host galaxy (optical) position is marked with '$+$' symbol.}
\label{nvss_C}
\end{figure*}

%------------------------------------------------------------------------------------------------------------------

\subsection{GRG5 (J044932-302638): A Giant Radio Quasar}
GRG5 has a strong core dominated feature along with a double lobe structure as seen in Figure~\ref{subfig-5}. The southern lobe is possibly contaminated by another unrelated radio source as it shows an apparent extension towards the western side. This object is also observed in the SUMSS which is slightly poorer in resolution and sensitivity when compared to NVSS. The spectroscopic redshift of $\sim$ 0.315 was obtained from \citet{2009MNRAS.399..683J} from which the projected linear size is $\sim 2.1$ Mpc. We infer the central black hole to be massive at 1.7$\times 10^{9}$ $M_{\odot}$ (Table~\ref{mass}). On the wise mid-IR colour-colour plots (Figure~\ref{ccplot}) GRG5 lies in the region of high excitation radio loud quasars which possibly makes it a giant radio quasar. The following properties further support the hypothesis of GRG5 being a giant radio quasar (GRQ): 
\begin{enumerate}
 \item Host galaxy of GRG5 shows excess emission in near and far UV (GALEX) \citet{galex}.
 \item It shows very high total and core (3.8$\times 10^{24}$ W$Hz^{-1}$) radio luminosities along with high infrared luminosity which is consistent with the nature of a quasar.
 \item The optical spectrum (Figure~\ref{spec5}) available in 6dF Galaxy Survey (6dFGS) (\citet{2009MNRAS.399..683J}) shows a
 very broad H$\beta$ line width ($\sim 2000$ km/s) which is commonly observed in quasars. A gaussian fit routine was used using IRAF (Image Reduction and Analysis Facility) and IDL (Interactive Data Language) on the lines of H$\beta$ and [OIII] doublet of the same spectrum to obtain the widths as shown in Figure~\ref{specgrq5}.
\end{enumerate}

%Start Table 2 ---------------------------------------------------------------------

\begin{table*}
\begin{center}
\begin{minipage}{130mm}
\caption{The J, H $\&$ K band magnitudes along with their respective errors are presented in this table. $M_{BH}$ is the mass of the black hole determined by using $M_{BH}$-$L_{K,bulge}$ 
correlation as given in Eq.\ref{Eq_MD17K}.}
\label{mass}
\begin{tabular}{ccccccccc}
\hline
GRG No. & J & J$_{Error}$ & H & H$_{Error}$ & K & K$_{Error}$ & $M_{BH}$ & $M_{BH-error}$ \\ 
       &   &      &   &      &   &      & ($10^{8} M_{\odot}$)    & ($10^{8} M_{\odot}$)  \\
\hline
1 & 17.31 & 0.23 & 16.25 & 0.17 & 15.64 & 0.2 & 13.8 & 2.4 \\ 
3 & 17.12 & 0.22 & 16.52 & 0.25 & 15.39 & 0.21 & 6 & 1.1 \\ 
4 & 16.62 & - & 16.58 & - & 14.79 & 0.13 & 34.6 & 3.8 \\ 
5 & 16.18 & 0.1 & 15.69 & 0.15 & 14.59 & 0.11 & 17 & 1.7 \\ 
7 & 17.03 & 0.23 & 16.02 & 0.19 & 15.35 & 0.19 & 6.3 & 1.0 \\ 
8 & 16.46 & 0.11 & 15.8 & 0.13 & 15.18 & 0.13 & 3.8 & 0.5 \\ 
9 & 14.32 & 0.08 & 13.69 & 0.11 & 13.11 & 0.07 & 5.8 & 0.4 \\ 
10 & 14.45 & 0.06 & 13.9 & 0.07 & 13.33 & 0.06 & 6.6 & 0.3 \\ 
13 & 14.14 & 0.05 & 13.54 & 0.06 & 13.1 & 0.06 & 5.9 & 0.3 \\ 
14 & 15.34 & 0.07 & 14.6 & 0.08 & 14.24 & 0.09 & 2.8 & 0.2 \\ 
% 15 & - & - & - & - & - & - & - &  \\ 
16 & 14.62 & 0.05 & 14.1 & 0.06 & 13.56 & 0.06 & 3.5 & 0.2 \\ 
17 & 14.19 & 0.05 & 13.48 & 0.06 & 12.98 & 0.05 & 5.0 & 0.2 \\ 
18 & 15.51 & 0.08 & 14.69 & 0.1 & 13.99 & 0.09 & 12 & 0.9 \\ 
% 19 & - & - & - & - & - & - & - & - \\ 
% 20 & - & - & - & - & - & - & - & - \\ 
% 21 & - & - & - & - & - & - & - & - \\ 
% 22 & - & - & - & - & - & - & - & - \\ 
23 & 14.18 & 0.07 & 13.28 & 0.08 & 12.97 & 0.08 & 3.1 & 0.2 \\ 
24 & 16.95 & 0.14 & 15.89 & 0.15 & 15.3 & 0.17 & 2.9 & 0.4 \\ 
% 25 & - & - & - & - & - & - & - & - \\ 
\hline
\end{tabular}
\end{minipage}
\end{center}
\end{table*}
% End Table 2 ---------------------------------------------------------------------

%Start Table 3 ---------------------------------------------------------------------

\begin{table}
\caption{The following table provides information about central velocity dispersion for hosts of GRGs whose data was available from SDSS. The velocity 
dispersion values were used to 
compute the black hole mass of the hosts of the following GRGs using $M_{BH}$-$\sigma$ relation.}
\begin{center}
\begin{tabular}{lcll}
\hline
No.  &    Velocity Dispersion & $M_{BH}$ \\ 
   &      (km$s^{-1}$)    & ($10^{9}$ $M_{\odot}$) \\
\hline   
GRG1  &  169.65 $\pm$ 35.02 &0.08 $\pm$	0.09 \\
GRG2  &  250.18	$\pm$ 48.54 &0.73 $\pm$	0.80 \\
GRG7  &  280.21	$\pm$ 30.28 &1.39 $\pm$	0.85 \\ 
GRG10 &  309.58	$\pm$ 07.41 &2.45 $\pm$	0.33 \\ 
GRG11 &  286.46 $\pm$ 70.63 &1.59 $\pm$  2.2 \\
GRG19 &  247.28	$\pm$ 07.35 &0.69 $\pm$  0.11 \\
GRG21 &  202.82	$\pm$ 14.13 &0.22 $\pm$  0.08 \\
\hline
\end{tabular}
\end{center}
\label{msigma-mbh}
\end{table}

%End Table 3 ---------------------------------------------------------------------

\vspace{-0.15in}
\subsection{GRG6 (J080248+492723)}
GRG6 is about 1.88 Mpc in projected linear size with its host galaxy at a redshift of $\sim$ 0.678. This is the most distant GRG in our sample. Not many GRGs of this size are known at higher redshifts. The FIRST map clearly resolves the radio core and lobes. The black hole mass of the host is unknown due to non-availability of K band and SDSS data. The FIRST map clearly resolves this GRG depicting its FR-II nature.

\subsection{GRG7 (J085701.76+013130.9)}
GRG7 has a very low surface brightness with an overall linear size of $\sim1.6$ Mpc. The FIRST map clearly resolves the radio core and the two radio lobes thus confirming its FR-II type nature.  Also, the host galaxy is located $\sim$ 3.5$\arcmin$ away from the GMBCG J134.30444+01.56066 galaxy cluster (which is at similar redshift as that of the GRG) as listed in the GMBCG catalog  of galaxy clusters \citep{2010ApJS..191..254H} .

\subsection{GRG8 (J105838.66+244535)}
GRG8 shows a peculiar morphology extending over 2.1 Mpc. The southern lobe is highly diffuse with an integrated flux  ($S_{i}$) of just $\sim$ 30 mJy at 1.4 GHz. In contrast to the southern lobe, the northern lobe shows a typical edge brightened lobe with an integrated flux ($S_{i}$) of $\sim$ 120 mJy.

\subsection{GRG9 (J105920-170920)}
GRG9 shows clear FR-II type morphology extending over 1 Mpc. The overall integrated flux ($S_{i}$) of GRG9 is only about 150 mJy at 1.4 GHz and has traces of diffuse emission.

\subsection{GRG10 (J132741+574943)}
The radio core of GRG10 shows an extension towards the southern lobe which could  be attributed to an unresolved radio jet. Interestingly, from Figure~\ref{subfig-10}, we observe diffuse emission beyond the hotspot of the northern lobe. The radio core as detected in FIRST coincides with the galaxy (SDSS J132741.32+574943.4) thus confirming it as a host. It is at a spectroscopic redshift of 0.1202 which makes its projected linear size to be 1.6 Mpc.
%From the SDSS DR12 spectrum we obtained the velocity dispersion ($\sigma$) of 309.58$\pm$7.41 km/s which is comparatively much  higher than the rest of the host  galaxies of present GRG sample. Using the $M_{BH}$-$\sigma$ relation  we obtained the mass of the central black hole to be 2.45$\pm0.33 \times 10^{9}$ $M_{\odot}$ as tabulated in Table~\ref{msigma-mbh}. We  derive a lower  black hole mass of 6.6$\pm0.3 \times 10^{8}$ $M_{\odot}$  using the $M_{BH}$-$L_{K,bulge}$ correlation as tabulated in Table~\ref{mass}. Based on the criteria mentioned in Section 2.2, we classify the host of GRG10 as LERG.

\subsection{GRG11 (J132743.50+174837)}
GRG11 is one of the more distant GRGs in our reporting sample with a redshift of $\sim$ 0.657. The FIRST map (Figure~\ref{subfig-11}) resolves GRG11 to be FR-II type with distinctive lobes.  The host galaxy appears highly reddened (Figure~\ref{optical}). We obtain a massive $\sim10^{9} M_{\odot}$ black hole mass via $M_{BH}$-$\sigma$, but it should be treated with caution due to high error, as seen in Table~\ref{msigma-mbh}.

\subsection{GRG12 (J200843+004918): An Extremely Large Radio Source?}
GRG12 has  angular size of 11$\arcmin$  showing a strong edge  brightened FR-II structure.  A faint galaxy SDSS J200843.37+004918.8 ($m_{r} = 20.76$) coincides with the faint radio core as seen in  Figure~\ref{subfig-12}, indicating it to be the host galaxy of GRG12. The host galaxy has a photometric redshift of  0.412 (SDSS-photoZ) which results in  $\sim$ 3.7 Mpc as its linear size. This would make GRG12 the largest GRG in the present sample as well as one of the largest  radio sources  known till date, after GRG J0422+1512 ($\sim$ 4 Mpc) \citep{2015AstBu..70...45A}, 3C236 ($\sim$ 4.6 Mpc) \citep{1974Natur.250..625W} and GRG J1420-0545, the biggest with a projected linear size of $\sim$ 4.9 Mpc. \citep{2008ApJ...679..149M}. The sizes of these GRGs (GRG J0422+1512, GRG 3C236 $\&$ GRG J1420-0545) were re-computed based on the cosmology adopted in this paper. Radio galaxies of such extreme physical sizes are extremely rare. GRG12 shows a clear FR-II  morphology and is quite similar to GRG J1420-0545. The mass of the central black hole could not be determined via $M_{BH}$-$L_{K,bulge}$ correlation as the host  galaxy is not detected  in 2MASS survey. It was also not detected in WISE survey in any bands, suggesting the galaxy to be quite far away. We need spectroscopic redshift data and detailed radio imaging observations to further probe this suspected extremely large radio galaxy.

% Start table 4----------------------------------------------------
\begin{table*}
\begin{center}
\begin{minipage}{170mm}
\caption{This table essentially shows mid-IR band magnitudes of GRGs. SNR is the profile-fit measurement signal-to-noise ratio in that band. This value is the ratio of the flux (of the band) to flux uncertainty in the W1 profile-fit photometry measurement. $10^{th}$ column shows the classification of GRGs using mid-IR WISE scheme, whereas the $11^{th}$ column shows GRG classification based on optical line flux ratios from SDSS spectrum. The WISE mid-IR luminosities ($L_{W1}$, $L_{W2}$, $L_{W3}$ \& $L_{W4}$) of GRGs in W1, W2, W3 and W4 bands as expressed in $L_{\odot}$.}
\label{wise}
\begin{tabular}{@{}ccccccccccccccc}
\hline
GRG No. & W1 & W1snr & W2 & W2snr & W3 & W3snr & W4 & W4snr & IR & Optical & $L_{W1}$ & $L_{W2}$  & $L_{W3}$  & $L_{W4}$ \\
               &        &            &      &            &      &            &      &            &      &             & ($10^{9}$) &($10^{9}$) & ($10^{10}$) & ($10^{11}$) \\
\hline
1 & 14.9 & 31.6 & 14.7 & 16.2 & 12.3 & 2.6 & 8.5 & 1 & LERG  &LERG  & 3.38 & 3.84 & 3.68 & 1.2 \\
2 & 14.9 & 31.9 & 14.8 & 16.5 & 12.4 & -2 & 8.9 & 0.1 & LERG  &  LERG & 6.51 & 7.37 & 6.62 & 1.6 \\ 
3 & 14.7 & 35.7 & 14.2 & 26.1 & 11.4 & 5.9 & 8.2 & 3.5 & LERG & - & 1.28 & 2.08 & 2.69 & 0.53 \\
4 & 12.6 & 45.5 & 11.2 & 51.7 & 8.3 & 41.1 & 5.9 & 24 & QSO & - & 32.98 & 122.18 & 169.27 & 15.89 \\  
5 & 13.8 & 42.4 & 12.5 & 47.5 & 9.6 & 30.3 & 7.2 & 12.7 &  QSO & QSO & 4.36 & 13.82 & 21.68 & 1.99 \\  
6 & 15.5 & 25.1 & 14.8 & 18.4 & 10.9 & 8.8 & 7.7 & 7.1 & LERG  & - & 5.82 & 10.03 & 36.4 & 7.68 \\ 
7 & 14.8 & 32.4 & 14.6 & 18.1 & 12.4 & 2.3 & 8.9 & -0.3 & LERG  & LERG & 1.27 & 1.53 & 1.19 & 0.27 \\ 
8 & 14.4  & 39.2 & 13.7 & 30.4 & 11.6 & 3.3 & 8.1 & 1.3  & HERG & - & 0.92 & 1.77 & 1.17 & 0.3 \\ 
9 & 12.2 & 47.2 & 12.1 & 47.6 & 11.3 & 6.6 & 8.7 & 0.5 & LERG & - & 1.59 & 1.71 & 0.35 & 0.04 \\ 
10 & 12.5 & 47.9 & 12.4 & 48.7 & 11.2 & 9.1 & 8.5 & 3.7 & LERG & LERG  & 1.62 & 1.86 & 0.56 & 0.07 \\ 
11 & 15.4 & 29.3 & 15.2 & 12.6 & 12.4 & 1.1 &9.1  & 0.6 & LERG & - & 5.65 & 6.66 & 9.08 & 1.87 \\
12 & 16.9 & 9.3 & 16.4 & 3.9 & 12.5 & -0.7 & 8.9 & -0.6 &  LERG &  - & 0.47 & 0.76 & 2.59 & 0.75 \\
13 & 12.3 & 47.8 & 12.3 & 42.1 & 11.5 & 4.7 & 8.5 & 0.5 &  LERG & - & 1.37 & 1.43 & 0.3 & 0.05 \\
14 & 13.6 & 44.1 & 13.4 & 38.6 & 11.5 & 6 & 8.8 & 0.4 &  LERG & -  & 0.57 & 0.67 & 0.41 & 0.05 \\
% 15 & - & - & - & - & - & - & - &  - & - & - & - & - & - & - \\ 
16 & 12.7 & 44.6 & 12.6 & 43.1 & 11.0 & 8.7 & 8.7 & 2.7 &  LERG &  - & 0.84 & 0.97 & 0.41 & 0.03 \\
17 & 12.5 & 46.9 & 12.3 & 48.7 & 10.6 & 12.6 & 8.9 & 0.5 & LERG &  - & 0.93 & 1.04 & 0.55 & 0.03 \\ 
18 & 13.4 & 38.6 & 13.2 & 29.9 & 11.6 & 1.1 & 7.8 & 1.8 & LERG & - & 2.58 & 3.16 & 1.31 & 0.45 \\ 
% 19 & - & - & - & - & - & - &  - & - & - &  HERG & - & - & - & - \\
20 & 14.9 & 31.5 & 13.9 & 25.8 & 11.0 & 7.2 & 8.1 & 1.4 &  QSO & - & 0.69 & 1.74 & 2.52 & 0.36 \\ 
21 & 14.5 & 35.5 & 13.9 & 30.5 & 11.4 & 7.3 & 8.6 & 3.4 & HERG & HERG & 1.41 & 2.38 & 2.47 & 0.33 \\ 
22 & 14.4 & 36.9 & 12.9 & 39.7 & 9.6 & 20.6 & 6.7 & 14.2 &  QSO & - & 10.07 & 37.02 & 81.36 & 11.19 \\  
23 & 12.1 & 36.8 & 12.0 & 38 & 11.1 & 7 & 8.5 & 0.4 & LERG &  - & 0.81 & 0.84 & 0.19 & 0.02 \\
24 & 14.7 & 32.6 & 14.2 & 19.6 & 11.3 & 5.2 & 8.0 & 1.5 & LERG & - & 0.57 & 0.94 & 1.3 & 0.27 \\ 
25 & 16.1 & 16.7 & 15.9 & 5.9 & 12.2 & 2.9 & 8.7 & 2.6 & LERG & - & 2.92 & 3.31 & 11.39 & 2.75 \\ 
\hline 
\end{tabular}
\end{minipage}
\end{center}
\end{table*}

% End table 4 ----------------------------------------------------------------------------------

%------------------------------------------------------------------------------------------------------------------
\begin{figure*}
%  \hspace{-0.6cm}
\includegraphics[height=3.72in,width=6.3in]{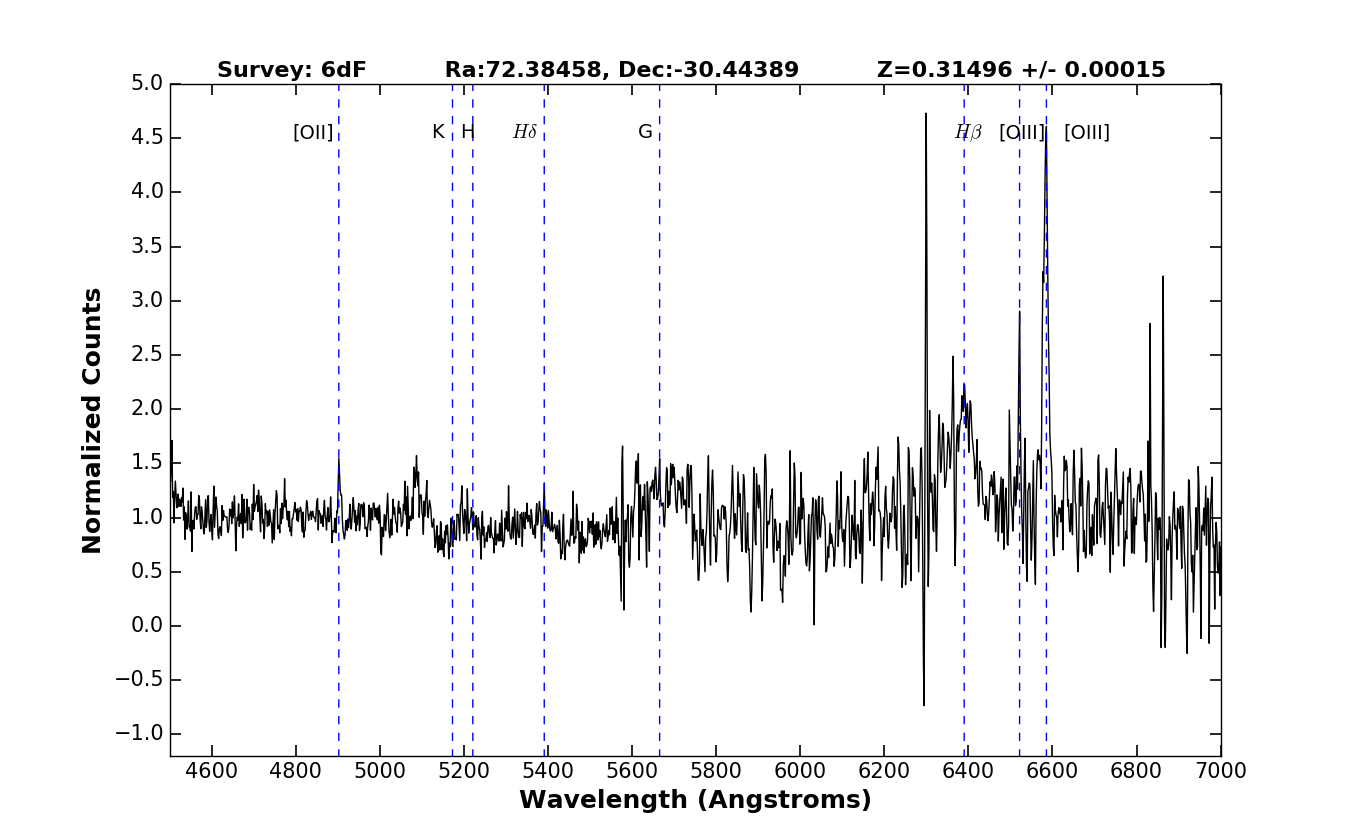}
\caption{Replotted optical spectrum of GRG5 using 6dFGS data.}
\label{spec5}
\end{figure*}

%------------------------------------------------------------------------------------------------------------------

\subsection{GRG13 (J203449-263036)}
As seen in Figure~\ref{subfig-13}, GRG13 does not display clear lobe hotspots. The emission around the core is more of diffuse nature. The radio core is very bright which coincides with the host galaxy (6dF J203449-263036) and has a spectroscopic redshift of 0.103 as obtained from \citet{2009MNRAS.399..683J}. GRG13 shows highly bent nature in the south-western lobe as compared to the north-eastern lobe. Due to presence of another strong radio source close to north-eastern lobe, the weak diffuse emission in this region is not very clearly seen in the NVSS map. From Figure~\ref{ccplot}, we observe GRG13 to lie in the LERG region.

\subsection{GRG14 (J205939+243423)}
GRG14 shows an asymmetric morphology as evidenced in Figure~\ref{subfig-14}. The southern lobe ($S_{i}$=58 mJy) shows a strong  hotspot surrounded by some extended diffuse emission and is closer to the radio core than the northern lobe ($S_{i}$=23 mJy). This asymmetry could be attributed to the existence of varied IGM near the southern lobe. The host galaxy, as seen in Figure~\ref{optical}, shows slight elongated morphology. Spectrum and spectroscopic redshifts are not available for the host (SDSS J205939.82+243423.9) of GRG14. However, SDSS photoZ  is about 0.116 which results in its projected linear size to be $\sim$ 1.1 Mpc. Based on Figure~\ref{ccplot} and a low black hole mass (Table~\ref{mass}), we infer GRG14 to be a LERG.

%---------------------------------------------------------------------------------------------------------------------
\begin{figure*}
\includegraphics[height=5.0in,width=6.5in]{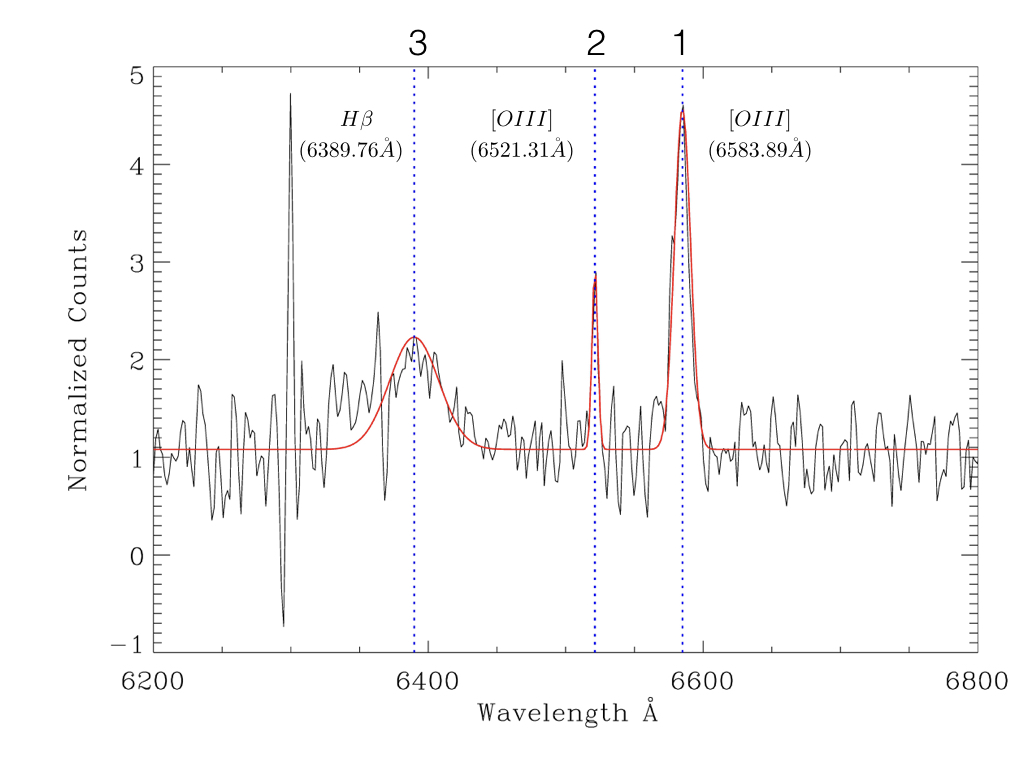}
\caption{The above figure shows zoomed in version of optical spectrum of host of GRG5 (J044932-302638) obtained from 6dFGS. A gaussian was
fit to spectral lines H$\beta$ and [OIII] doublet to obtain  the widths. The widths are as
follows: $ (1) [OIII]:\bigtriangleup \lambda=13.77 $\AA$, \bigtriangleup $v$ = 627.3 km/s, (2) [OIII]: \bigtriangleup \lambda=4.91 $\AA$ , \bigtriangleup $v$ = 226.3 km/s , (3) H\beta: \bigtriangleup \lambda=42.39 $\AA$, \bigtriangleup $v$ = 1990.2 km/s $. The broad nature of H$\beta$ line is clear indicator of its quasar nature.}
\label{specgrq5}
\end{figure*}

%---------------------------------------------------------------------------------------------------------------------

\subsection{GRG15 (J223301+131502): A GRG hosted by a BCG}
The radio morphology of this GRG is non-linear and peculiar. We observe the host galaxy SDSS J223301.30+131502.5 in optical band (Figure~\ref{optical}) to coincide with the radio core whose position is marked  $'+'$ on the NVSS colour contour map (Figure~\ref{subfig-15}). On the western side of the core, we observe a  bright unresolved peak  which we conjecture to be a  bright knot structure on  the jet. A similar but fainter knot is also observed on the eastern jet  closer to the radio core. Beyond these two close hotspot/knots on either side of the radio core are the outer lobes  of GRG15 which appear  quite diffuse. The distance between  
outer lobes is measured to be $\sim$ 16$\arcmin$, while the host galaxy has a photometric redshift of 0.093 which makes the projected linear size of the source as 1.71 Mpc. No FIRST data is available for this source therefore a higher resolution image  is required to extract a clear picture of its fairly complicated morphology.

\begin{figure}
 \includegraphics[height=3.2in,width=3.2in]{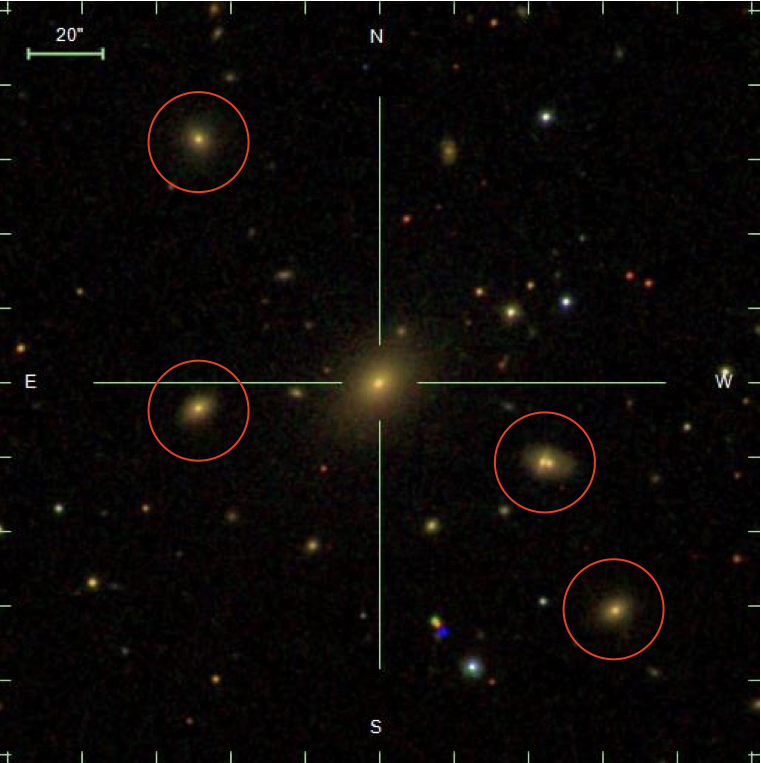}
 \caption{3.3' $\times$ 3.3' optical SDSS image of environment of GRG15. Red circles are other galaxies of same cluster and at the center is the host galaxy of GRG15.}
 \label{env}
\end{figure}

Interestingly, in the GMBCG catalog  of galaxy clusters \citep{2010ApJS..191..254H}, we find the host of GRG15 listed as the brightest cluster galaxy (BCG) in cluster GMBCG J338.25544+13.25070. This is a bright cD galaxy  (SDSS $m_{r} = 15.21$) near the cluster center. The host is also listed as BCG in maxBCG galaxy cluster catalog \citep{2007ApJ...660..239K} and the WHL galaxy cluster catalog \citep{2012ApJS..199...34W}. The SDSS shows the redshift of the  cluster is near to 0.102, close to the photometric redshift (0.093) of host cD galaxy. Having a peculiar radio morphology and residing in a dense cluster environment, we find GRG15 to be very unusual  and  thus  important for  understanding the evolution of GRGs in dense cluster environments. Further studies are required to understand the physical  and  spectral evolution of this galaxy which has attained  projected linear size of 1.71 Mpc inspite being in a dense cluster environment (Figure~\ref{env}). This GRG challenges  one of the common beliefs that GRGs grow to their gigantic sizes due to their preferential location in the sparse galactic environment. This GRG also demonstrates that environment alone may not play a primary role in growth of the GRGs, as commonly believed in earlier works. Rather, a powerful AGN is at work here. While it is well known that BCGs preferentially have radio loud AGN \citep{1994JApA...15..275B, 2007MNRAS.379..894B}, observations so far imply that very small fraction of BCGs have large scale radio structures. This result  highlights  the need for more detailed studies of the external environments for other GRGs as well as the physical conditions for  the BHs  powering these massive objects such as their mass, spin and mass accretion rate. This can be achieved via future high resolution X-ray, optical and IR spectroscopic studies of the host galaxies of these GRGs.

%Start Table 5 ---------------------------------------------------------------------
% 
% \begin{table}
% \begin{center}
% \caption{WISE mid-IR luminosities($L_{W1}$,$L_{W2}$,$L_{W3}$ \& $L_{W4}$) of GRGs in W1, W2, W3 and W4 band expressed in $L_{\odot}$.}
% \label{luminosity}
% \begin{tabular}{ccccc}
% \hline
% No. & $L_{W1}$ & $L_{W2}$  & $L_{W3}$  & $L_{W4}$ \\
%   & ($10^{9}$) &($10^{9}$) & ($10^{10}$) & ($10^{11}$) \\
% \hline
% 1 & 3.38 & 3.84 & 3.68 & 1.2 \\ 
% 2 & 6.51 & 7.37 & 6.62 & 1.6 \\ 
% 3 & 1.28 & 2.08 & 2.69 & 0.53 \\ 
% 4 & 32.98 & 122.18 & 169.27 & 15.89 \\ 
% 5 & 4.36 & 13.82 & 21.68 & 1.99 \\ 
% 6 & 5.82 & 10.03 & 36.4 & 7.68 \\ 
% 7 & 1.27 & 1.53 & 1.19 & 0.27 \\ 
% 8 & 0.92 & 1.77 & 1.17 & 0.3 \\ 
% 9 & 1.59 & 1.71 & 0.35 & 0.04 \\ 
% 10 & 1.62 & 1.86 & 0.56 & 0.07 \\ 
% 11 & 5.65 & 6.66 & 9.08 & 1.87 \\
% 12 & 0.47 & 0.76 & 2.59 & 0.75 \\ 
% 13 & 1.37 & 1.43 & 0.3 & 0.05 \\ 
% 14 & 0.57 & 0.67 & 0.41 & 0.05 \\ 
% % 15 & - & - & - & - \\ 
% 16 & 0.84 & 0.97 & 0.41 & 0.03 \\ 
% 17 & 0.93 & 1.04 & 0.55 & 0.03 \\ 
% 18 & 2.58 & 3.16 & 1.31 & 0.45 \\ 
% % 19 & - & - & - & - \\ 
% 20 & 0.69 & 1.74 & 2.52 & 0.36 \\ 
% 21 & 1.41 & 2.38 & 2.47 & 0.33 \\ 
% 22 & 10.07 & 37.02 & 81.36 & 11.19 \\ 
% 23 & 0.81 & 0.84 & 0.19 & 0.02 \\ 
% 24 & 0.57 & 0.94 & 1.3 & 0.27 \\ 
% 25 & 2.92 & 3.31 & 11.39 & 2.75 \\ 
% \hline
% \end{tabular}
% \end{center}
% \end{table}

%End Table 5---------------------------------------------------------------------

\vspace{-0.2in}

\subsection{GRG16 (J225039.15+284445.5)}
The radio core coincides with the optical host (SDSS J225039.15+284445.5) perfectly which is at a redshift of $\sim$ 0.097 thus making its projected linear size as 0.9 Mpc. In Figure~\ref{subfig-16}, we see the evidence of the source showing hotspots which are asymmetric. We also observe diffuse emission beyond the hotspots.

\subsection{GRG17 (J225615.1-361759)}
GRG17 shows highly diffuse nature with its southern lobe more diffuse than the northern one (Figure~\ref{subfig-17}). The northern lobe has some hint of a hotspot while the southern one is highly diffuse and has a very low surface brightness. The redshift as obtained from \citet{2009MNRAS.399..683J}, is 0.09 which makes its projected linear size as 1.5 Mpc. Only the core is significantly observed in the SUMSS.

\subsection{GRG18 (J230444-105048)}
GRG18 has  low surface brightness and does not show clear hotspots, as seen from Figure~\ref{subfig-18}. Its radio core is seen to coincide with optical host (2MASX J23044483-1050474). The redshift, as obtained from \citet{2009MNRAS.399..683J}, is about 0.21 which results in its linear projected size to be $\sim$ 0.9 Mpc. With no clear lobes or hotspots observed, it is more likely to be of FR-I nature.

\subsection{GRG19 (J231201.27+135655.9)}
The radio core is seen to coincide with the optical host (SDSS J231201.27+135655.9) and is at a redshift of $\sim$ 0.140 which results in its projected linear size of $\sim$ 1.7 Mpc. In Figure~\ref{subfig-19}, we observe the southern lobe to be brighter than the northern lobe. The northern lobe shows a hotspot towards the far end. Based on the criteria mentioned in Section 2.2, we classify the host of GRG19 as LERG.

\subsection{GRG20 (J231620-010207)}
In Figure~\ref{subfig-20}, we see the distinct lobes of GRG20 with the radio core off centered towards the eastern lobe. The radio core, as seen in FIRST, coincides with the optical host (SDSS J231620.15-010207.3). At a redshift of 0.221 (SDSS photoZ) gives its projected linear size is 1.5 Mpc. Through the WISE colour-colour diagram we observe it to lie in the QSO region. Further optical studies are required to confirm the true nature of the host galaxy.

\hspace{-0.16in}
\subsection{GRG21 (J232623+245840)}
Having an extremely large size of 2.9 Mpc, GRG21 is the third largest in our reporting sample. We obtained the spectroscopic redshift (0.2549) and velocity dispersion ($\sigma$ = 202.82$\pm$14.14 km/s) from SDSS DR12 spectrum. Using the $M_{BH}$-$\sigma$ relation we computed the mass (Table~\ref{msigma-mbh}) of the central black hole to be 0.22$\pm0.08 \times 10^{9}$ $M_{\odot}$. It has a high core radio luminosity ($L_{c}$) $=$ 2$\times 10^{25}$ W$Hz^{-1}$ computed from the NVSS maps.  Based on the criteria  mentioned in Section 2.2  and  mid-IR colour-colour (Figure~\ref{ccplot}) plot, we classify it to be a HERG. GRG21 is highly core-dominated radio galaxy having a bright double structure with the southern lobe (having a hotspot) being brighter ($S_{i}$=89 mJy) than the northern lobe ($S_{i}$=44 mJy), as seen in Figure~\ref{subfig-21}. 

\subsection{GRG22 (J232849.99-082512.7)}
At a redshift $\sim$ 0.555 its measured projected linear size is 1.9 Mpc, where the host galaxy (SDSS J232849.99-082512.7) coincides with the radio core perfectly. From  Figure~\ref{subfig-22} we observe that GRG 22 is highly asymmetric and the radio core lies towards the northern lobe. The core is clearly resolved in FIRST and can be seen from the black contours plotted in Figure~\ref{subfig-22}. The black contours seen in  Figure~\ref{subfig-22} are from FIRST where the core is clearly resolved. This GRG is also mentioned in \citet{2015salt.confE..34B} where they have obtained the spectroscopic redshift of the host galaxy which is lesser than that of SDSS as quoted above. Taking \citet{2015salt.confE..34B} redshift for the host J232849.99-082512.7 as $0.3839\pm0.0073$ we obtain slightly lesser projected linear size $\sim$ 1.5 Mpc and $P_{1.4GHz}$  $\sim$ $1.8\times10^{25} WHz^{-1}$. 

\subsection{GRG23 (J233552.1+521539)}
The host 2MASX J233552.14+521538.7 coincides with the radio core and is observed to be towards the north-eastern lobe. With a redshift of $\sim$ 0.07 \citep{2012ApJS..199...26H}, its projected linear size measures 0.85 Mpc. From Figure~\ref{subfig-23}, we observe GRG23 to be an asymmetric GRG with the south-western lobe being very elongated and showing a hotspot. But there is no indication of hotspot in the north-eastern lobe and the jets terminates abruptly without a lobe or hotspot.

\subsection{GRG24 (J234929-000305)}
The host galaxy (SDSS J234929.77-000305.8) for GRG24 is seen to coincide with the radio core which is prominently seen in FIRST. With a redshift of $\sim$ 0.187 its projected linear size measures 0.84 Mpc. From Figure~\ref{subfig-24}, we observe the eastern lobe to show further diffuse emission while the western lobe is elongated. The black contours seen in Figure~\ref{subfig-24} are from FIRST where the core is clearly resolved.

\subsection{GRG25 (J235531+025607)}
GRG25 is the second largest GRG and one of the more distant from our reporting sample. The SDSS photoZ is of redshift $\sim$ 0.657 thus giving the projected linear size as $\sim$ 3.5 Mpc. There are scarcely any GRGs known at this redshift with sizes greater than  $\sim$  2 Mpc. We observe the host galaxy (SDSS J235531.63+025607.1) to coincide with the radio core. This GRG shows a double lobed morphology, as evidenced from Figure~\ref{subfig-25}. The black contours as seen in Figure~\ref{subfig-25} are from FIRST where the core is clearly resolved.

%----------------------------------------------------------
\begin{figure}
\hspace{-0.8cm}
\includegraphics[height=4.7in,width=4.15in]{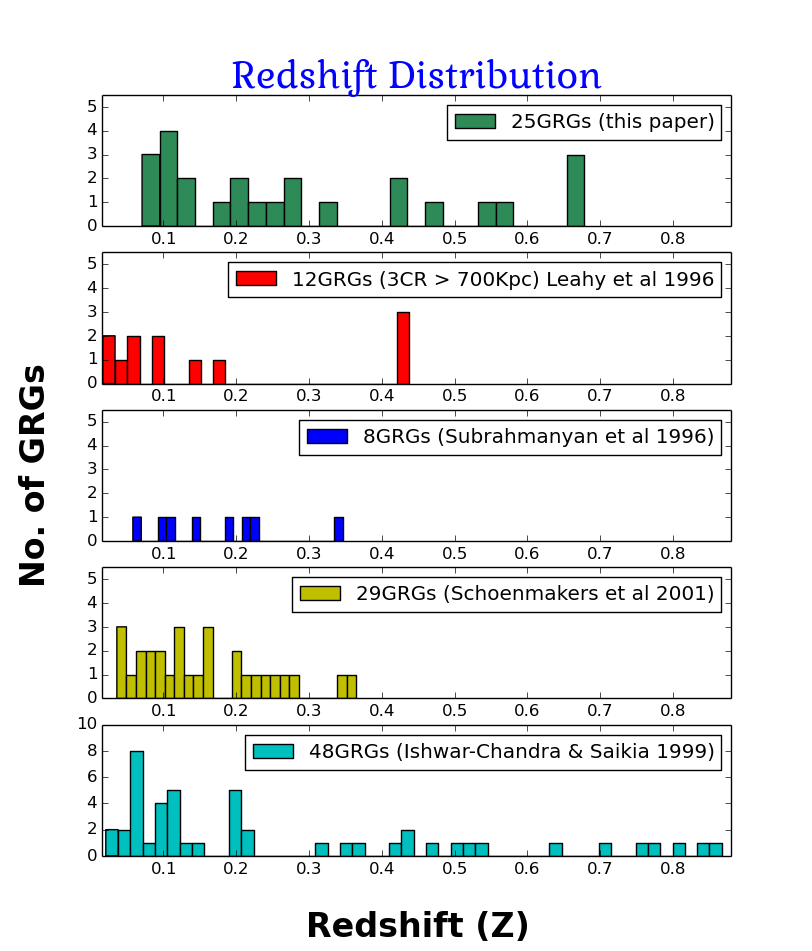}
\caption{A comparison of redshift distribution of GRGs of linear sizes $\geq$ 700 Kpc, both
from our  present sample as well as from other samples of GRGs in the literature.}
\label{barplot}
\end{figure}
%----------------------------------------------------------

\vspace{-0.1in}
\section{Discussion }
\subsection{Black hole mass of GRG}
It is commonly observed that megaparsec scale radio galaxies are hosted by bright elliptical galaxies, with a few rare exceptions recently discovered where the host galaxies are clearly spirals \citep{2011MNRAS.417L..36H,2014ApJ...788..174B}. It is conjectured that GRG  hosts contain super massive black holes of masses  $10^{8}-10^{9} M_{\odot}$ in their galactic centres  which are responsible for powering their large-scale jets and  terminal  lobes \citep{1984RvMP...56..255B}. It still remains a mystery as to how this system of jets and lobes grew  to their current  gigantic sizes. Our knowledge about the  physics of black hole  growth in AGN, and the impact of  the energetic relativistic jets on the surrounding medium is very limited and not yet fully understood. Several papers have pointed out (e.g. \citet{2000ApJ...543L.111L,2004MNRAS.351..347M,2007ApJ...658..815S}), that AGNs harboring  black holes of masses above a few $\times  10^{8}$ $M_{\odot}$ in elliptical galaxies only are capable of launching large scale ($> 100$~kpc) radio jets.  

In our present sample, almost all the GRGs are hosted by elliptical galaxies and we show that almost all of them host SMBHs (super massive black holes) of mass range $ 10^{8}$-$10^{9}$ $M_{\odot}$ as computed in Table ~\ref{mass} and ~\ref{msigma-mbh}.  The mass of SMBH of all the GRGs except GRG2, GRG10, GRG11, GRG12, GRG15, GRG19 to GRG22  and GRG25 was computed using the $M_{BH}$-$L_{K,bulge}$ correlation \citep{2007MNRAS.379..711G} as seen in Table~\ref{mass}. Interestingly, none of the black hole masses estimated by us fall below  $10^{8}$ $M_{\odot}$ thus supporting the earlier results suggesting that there is possibly a minimum  BH mass that separates  radio-loud  galaxies from radio-quiet ones. Amongst our sample of 25, the mass of central BH of  GRG1, GRG4, GRG5 and GRG18 (Table~\ref{mass}) is exceptionally high, all above $10^{9}$ $M_{\odot}$.

%-------------------------------------------------------------------------------------------
%-------------------------------------------------------------------------------------------

\vspace{-0.18in}

\subsection{Demographics of GRGs} 
From literature, we know that very few GRGs are known till date ($\sim$300) relative to normal radio galaxies and even a small subset of GRGs are known to be at high redshifts ($z > 0.8$). The effect of inverse compton losses on the ageing of the particles in radio lobes rapidly increases with redshift as the energy density of the cosmic microwave background increases as (1+z)$^{4}$ \citep{1993ppc..book.....P} which when combined with adiabatic expansion and synchrotron losses diminishes the luminosity of an object rapidly thus making it difficult to be detected in flux limited surveys.
Therefore, it indicates that the detected high redshift GRGs are intrinsically very powerful. This is reflected in our sample of 25 where we have seven GRGs having their redshift $\geq$ 0.3. In Figure~\ref{barplot}, we can see a comparison of redshift distributions of our GRGs (25) with other samples of known GRGs. These samples of known GRGs \citep{1996IAUS..175..157L,1996MNRAS.279..257S,1999MNRAS.309..100I,2001A&A...374..861S} were chosen for uniformity and their characteristics, like linear sizes and redshift, were recomputed and reconfirmed according to the cosmology adopted in this paper. Our sample of 25 GRGs clearly show a wide range of redshifts from 0.07 to $\sim$ 0.68. Due to  the faint nature of  GRGs it is extremely difficult to detect such large sized distant sources at much higher redshifts ($z > 0.5$) and needs high sensitivity deep images at  fairly low radio frequencies ($\nu < 100$ MHz) with sufficiently high resolutions where these extended lobe structures shine more prominently. Upcoming low frequency surveys from uGMRT \footnote{\url{http://www.gmrt.ncra.tifr.res.in/gmrt_hpage/Upgrade/index.html}}, LOFAR \citep{lofar}, MWA \citep{mwa} and future surveys like SKA-LOW will be able to discover many GRGs at very high redshifts. Increasing the number of such more rare distant GRGs will help us constrain the evolutionary models in cosmic time. 

Based on the intensity ratios of the optical emission lines, radio-loud AGN can be classified in two distinct populations namely the low-excitation radio galaxies (LERG) and high-excitation radio galaxies (HERG). \citet{2012MNRAS.421.1569B} differentiated between the LERG and the HERG by the efficiency of matter being accreted onto the central black hole. It is observed that LERGs are radiatively inefficient accretors with most of the energy from the accretion being channeled into the radio jet. On the other hand, HERGs are radiatively efficient and show strong evidence of  nuclear activity in the optical band. LERGs show high black hole mass along with low accretion rate which is exactly opposite for HERGs.
It is clear that jets can be generated both with and without a radiatively efficient disc. As seen from Table~\ref{wise} (11th column - 'optical'), we have classified 6 GRGs using SDSS line flux ratios and equivalent widths into HERGs and LERGs. We do not observe any trend in the hosts of the GRGs being preferentially high excitation or low excitation. Such a classification of large sample of GRGs will perhaps reveal the underlying dichotomy or any other trends for GRGs.

\subsection{WISE mid-IR studies of GRGs}
\begin{figure*}
\hbox{
\hspace{-0.6cm}
\includegraphics[height=4.2in,width=4.21in]{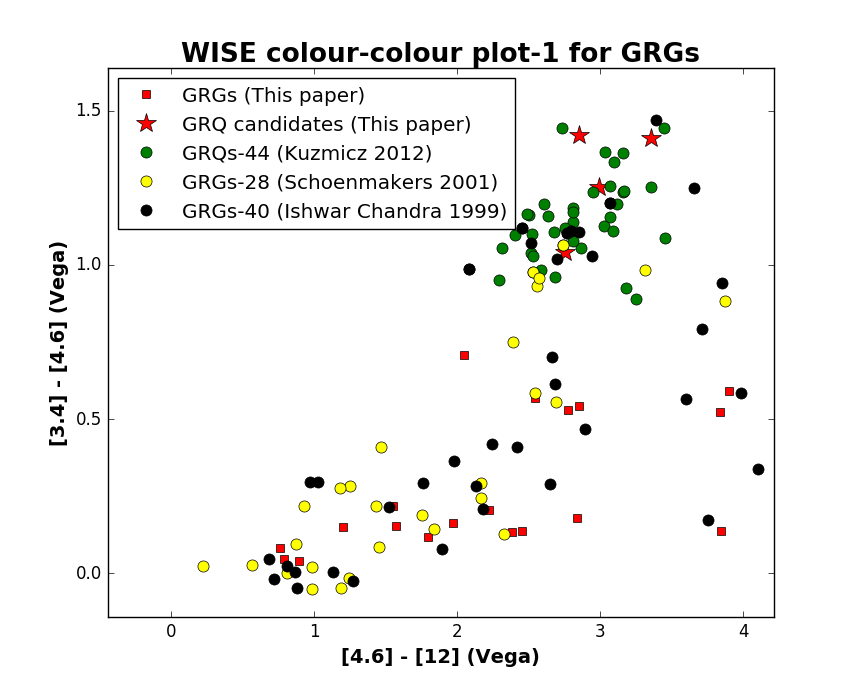}\hspace{-1.1cm}
\includegraphics[height=4.2in,width=4.21in]{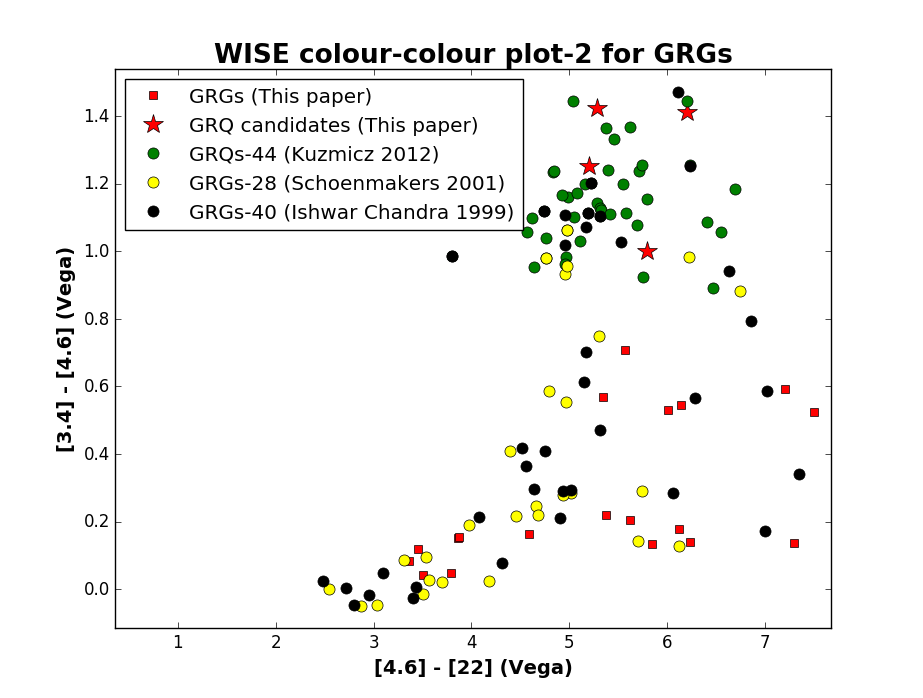}
}
\caption{The above plots illustrate where our GRGs lie on the mid-IR colour-colour plots made using WISE mid-IR
magnitudes W1, W2, W3 and W4 (3.4, 4.6, 12, 22-$\mu$m Vega magnitudes). Left plot: The region of star-forming galaxies is
[W2] - [W3] $>$ 2 \citep{2010AJ....140.1868W} and LERGs are concentrated mostly in the region: [W1]-[W2] $<0.8$ and [W2]-[W3] $<2.5$.
Right plot: The x-axis here is [W2] - [W4]. \citet{2014MNRAS.438.1149G} have shown that in both plots, quasars and BLRGs have similar colours.}
\label{ccplot}
\end{figure*}

\citet{2014MNRAS.438.1149G} differentiated between LERG, HERG, NLRG, BLRG and radio loud quasars with the help of WISE (mid-IR) colour-colour plots using the Cambridge radio surveys (3C, 4C etc). We place our GRGs on the WISE mid-IR colour-colour plots (Figure~\ref{ccplot}) to classify their AGN using the four mid-infrared bands. To our knowledge this kind of  classification is being done for the first time for GRGs. \citet{2014MNRAS.438.1149G} have shown that LERGs have redder colours and lie at the bottom-left region of the WISE colour-colour plots.
From Figure~\ref{ccplot}, we observe that majority of the GRGs in our reporting sample lie in the LERG region ([W1] - [W2] $<$ 0.8).
In Figure~\ref{ccplot}, we plot the excitation type of nearly 100 AGN of GRGs and it is evident from this plot that GRGs do not show a
preference towards high excitation or low excitation types. However the GRQs are all of high excitation type (see discussion below).

%---------------------------------------------------------------------------------------------------------------------------------------------------------------------------------
%-------------------------------------------------------------------------------------------------------------------------------------------------------------------------------------

\vspace{-0.15in}
\subsubsection{An efficient method for quasar identification}
From standard AGN models we know that optical or UV emission obscured by dusty structures around the accretion disc is re-radiated in the mid-IR. Thus, hidden quasars can be found by means of mid-IR observations such as WISE. \citet{2014MNRAS.438.1149G} have shown that quasars have higher mid-IR luminosities compared to other AGN types and it is also observed to be holding for vast range of redshifts. The re-radiated optical or UV emission from AGN dominates in 22$\mu$m flux (W4). \citet{2012AJ....144...49W} have also shown using SDSS quasars that WISE bands are efficient in finding quasars. From \citet{2014MNRAS.438.1149G} colour-colour plots, we observed that radio loud quasars cluster in one particular region of the plots defined by [W1] - [W2] $>$ 0.85 and [W2] - [W3] $>$ 2. 
We further validated it by putting known GRQs (44) from \cite{2012MNRAS.426..851K} on the colour-colour plot and find that all the GRQs lie in the QSO region of the plot as mentioned above. Further, we observe that GRG4, GRG5, GRG20 and GRG22 from our new sample lie in QSO region of the plots (Figure~\ref{ccplot}) which indicates that they are possibly radio loud quasars. 
Radio loud quasars are known to have a very high core radio luminosity as well as high infrared luminosity which is indeed very well exhibited by GRG4 and GRG5 of our reporting sample. They also show high black hole mass ($> 10^{9}$ $M_{\odot}$) which is another indicator of their quasar nature.
This claim is further validated by the optical spectrum of GRG5 (Figures ~\ref{spec5} $\&$~\ref{specgrq5}). GRQs B0750+434 $\sim$ 2.45 Mpc \citep{2001A&A...374..861S} $\&$ HE1127-1304 $\sim$ 2.1 Mpc \citep{1998MNRAS.299L..25B} are two of the largest GRQs known till date. This places GRG4 and GRG5 (also 2.4 Mpc and 2.1 Mpc respectively) from our sample amongst the largest known GRQs. These giant radio quasars are very important for the ongoing studies on AGN unification schemes. Therefore, we show that in absence of an optical spectrum, the WISE mid-IR colour-colour plots along with the radio and infrared luminosities provide a good alternative method to identify GRQs and classify GRGs into different AGN types.

\vspace{-0.17in}
\subsection{P-D Diagram}
\begin{figure}
\includegraphics[height=3.7in,width=3.8in]{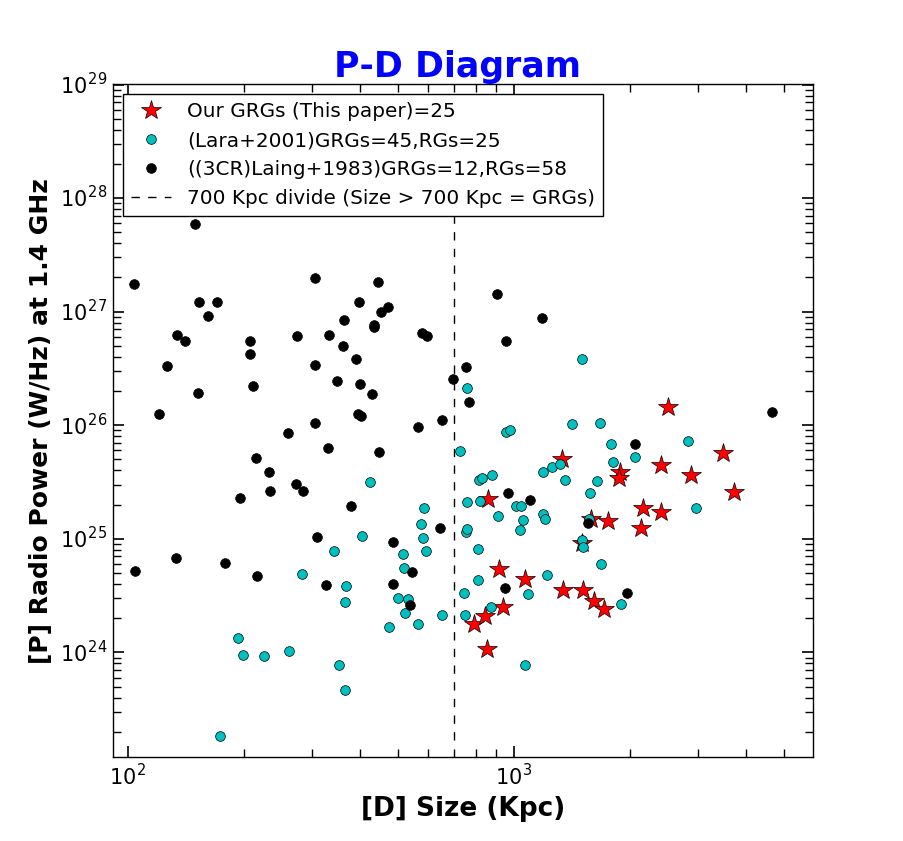}
\caption{Above figure shows the newly discovered 25 GRGs plotted on the P-D diagram along with other known RGs and GRGs from different papers. Sizes and powers of known RGs and GRGs were 
recomputed for uniformity.}
\label{pd-diag}
\end{figure}

Traditionally, a P-D diagram \citep{1982IAUS...97...21B} is a plot between radio power (P) at some fixed frequency and the linear size (D) of radio galaxies. Using tools such as the P-D diagram,  we can study the evolution of radio sources \citep{1997MNRAS.292..723K,1999MNRAS.309..100I,2002A&A...391..127M,2004AcA....54..249M}.
In Figure~\ref{pd-diag}, we plot the P-D diagram for our new sample GRGs (25) along with a sample of known GRGs from literature \citep{1983MNRAS.204..151L,2001A&A...370..409L}. This is one of the biggest (82) sample (new and previously known) of GRGs ever to be compared in a P-D diagram. For sources from 3CR\footnote{\url{http://www.jb.man.ac.uk/atlas/}}, flux at 1.4 GHz was obtained using the given spectral index. This enabled us to compute the radio power at 1.4 GHz using Eq~\ref{power}. The 3CR sample has a total of 85 sources (DRAGNs : Double Radio source Associated with Galactic Nucleus) of which only 70 were above 100 Kpc in total size. Only these 70 sources were plotted on the P-D diagram. Based on our definition ($D\geq700Kpc$) of GRGs, only 12 sources are giants from the 3CR sample of 70 radio galaxies (RGs). The second chosen sample for comparison study in this plot is from \cite{2001A&A...370..409L} where the total number of sources are 70 (45GRGs $\&$ 25RGs). The sample in \cite{2001A&A...370..409L} is also constructed from NVSS (similar to our sample) and it provided us the 1.4 GHz flux in mJy with total source size in arcminutes. The linear sizes and radio power were recomputed for all the objects (on the plot) using the updated cosmological parameters (as stated in section 1) so as to maintain uniformity and avoid biases in comparison.

It is well evident from the P-D diagram that large sized radio sources tend to be less radio luminous at 1.4 GHz frequency. This perhaps indicates that as these radio galaxies grow in size (megaparsec scales), they become less luminous. The diminishing luminosity is possibly due to adiabatic expansion and radiative energy losses. In some cases, it might also be because the radio-loud phase is  turned off.  Based on the evolutionary models by \citet{1999Natur.399..330B}, after an initial rapid increment, the luminosity of the radio lobes decreases rapidly as the source grows to a larger extent. The sources evolve towards lower power and larger size as they age. This is suggested by the absence  of radio sources  in the upper right corner of the P-D diagram (large sizes and high powers).
Assuming that the sample used is a reasonable representation of the population, we support our claim through the results of the Mann-Whitney U test \citep{mann1947test} which is a non-parametric test for equality of sample medians. Galaxy size of 700 kpc is used to segregate the sample into two - those lower than 700 kpc are refered as RGs while the others as GRGs. We test the null hypothesis of the median power of RGs and GRGs being equal against the alternative hypothesis of radio power of GRGs being lower than that of RGs. \\
%$H_{0}: median (P_{1.4GHz} (GRGs)) = median(P_{1.4GHz} (RGs))$ \\
%$H_{1}: median (P_{1.4GHz} (GRGs)) < median(P_{1.4GHz} (RGs))$ \\

With the test statistic $W = 2779$ and a p-value = 0.0277, we have sufficient evidence to reject the null hypothesis.

Interestingly, no sources are seen in the extreme lower right corner of the P-D diagram either, i.e. of  very large sizes and  very low powers. The existing surveys which have discovered most of the known GRGs so far are unable to detect these extreme GRGs due to sensitivity limitation of the surveys. The models predict that these yet unobserved sources are possibly the final stage of evolution of a radio galaxy before it fades away. Therefore, future low frequency radio surveys \citep{2015aska.confE.173K} (below 100 MHz) with LOFAR, SKA-low etc. will be able to uncover the hitherto undetected population of giant radio sources in the bottom right part of the P-D diagram. These sources  will provide vital information on the  evolution of radio galaxies to such extreme limits and on duty cycle of accreting massive black holes in host galaxies.

Such studies of GRGs are of vital  importance  for making progress in understanding the unusual nature of these extreme galaxies. The multi-frequency observations will allow us to obtain sensitive maps of the low surface brightness diffuse emission, spectral index maps, break frequency, and  spectral ages of GRGs. Under project \textit{SAGAN}, reporting and analysis of more new GRGs from our discovery sample will form part of our next paper in this series which is under preparation currently. Some very interesting results from multi-wavelength analysis of  larger sample of GRGs shall be discussed in our consecutive papers of this same series. Recent unique work carried by others like \citet{2016ApJS..224...18P} will be quite useful for forming a bigger and complete sample of GRGs after supplementing it with existing and new optical data. We would like to highlight that all the studies carried out in the present paper were done using the archival data from various surveys in radio, optical and infrared bands.

%----------------------------------------------------------
%----------------------------------------------------------

\vspace{-0.2in}
\section{Conclusions and Summary}
In this paper, we present the discovery of 25 new GRGs from the NVSS survey and their multiwavelength studies. This reporting sample of GRGs is not a complete sample of search in NVSS. Efforts to make a complete sample from these searches in NVSS is under progress and will be presented in the forthcoming papers. The NVSS was completed  more than 15 years ago and these new discoveries signify its continued importance and confirm that it is far from becoming obsolete. Large size and relatively low surface brightness of GRGs makes it difficult to detect them via automated detection in various radio surveys. Combining NVSS images with optical data from SDSS as well as archival  data from FIRST and SUMSS, we provide a complete study of the morphology of these sources. Further, using WISE  colours, we could classify the hosts of GRGs into different accretion types (HERGs, LERGs, etc.). Using 2MASS K band and SDSS spectroscopic data, we estimate the mass of the central black hole in  $\sim$ 50 $\%$ of hosts of these GRGs. The estimated black hole masses are in the range of $10^{8} - 10^{9}$ $M\odot$. \\
A brief summary of our results is as follows:
\begin{enumerate}
\item Considering that only $\sim$ 300 GRGs are known till date, our contribution of 25 new GRGs enhances the sample by $\sim$ 9$\%$. Furthermore, all the GRGs in our  sample have radio power above 10$^{24}$ W Hz$^{-1}$ at 1.4 GHz.

\item Optical spectra for 13 GRGs were available from the archive. Almost all the optical spectra presented in this paper show detection of H$\alpha$, H$\beta$, [O II], [O III], NII, SII thus confirming their AGN nature. We also classified a few GRGs into HERGs and LERGs using the SDSS line flux ratios.

\item We proposed a new method for identifying radio loud quasars without an optical spectrum. This was done by  estimating its mid-IR luminosity from WISE, radio  luminosity and its location on the WISE colour-colour plots. We find that the radio loud quasars cluster in one particular region of the plots defined by [W1] - [W2] $>$ 0.85 and [W2] - [W3] $>$ 2. We find that GRG4, GRG5, GRG20 and GRG22  show very high radio core luminosities as well as high mid-IR  luminosities which are strong indications of their quasar nature. This was further confirmed  with the optical spectrum of GRG5.

\item We show, for the first time, the classification of GRGs into QSOs, LERGs, HERGs, etc. using the WISE mid-IR colour-colour plots. From this classification, we notice that nearly half of GRGs in our reporting sample are LERGs. 

\item None of the host galaxies of GRGs in our sample fall on the star forming (SF) region ([W2]-[W3] $\geq 2.5$ and [W1]-[W2] $\leq$ 1) of the WISE mid-IR colour-colour plots, from which we infer that host galaxies of GRGs are mostly passive, red ellipticals without significant star-forming activity at the present moment. However, more studies are required to investigate this with a larger sample.

\item GRG15  clearly challenges the common belief that GRGs grow in sparse, non-cluster environment. The optical observation of the environment of this GRG host clearly shows that it is a cD-like brightest cluster galaxy which lies in a dense central region of the cluster. This shows that the environment alone does not play a major role in their exceptional large sizes.  

\item The P-D diagram suggests that GRGs ($>$0.7 Mpc), in comparison to normal radio galaxies, are less luminous and less powerful, indicating the end stage (de-energising) of radio galaxy life-time. Moreover, we observe a lack of extremely large and powerful sources in the top right corner of the P-D diagram.

\item Also, the P-D diagram suggests a cutoff in the GRG radio power at $\sim$ $10^{26}$ W$Hz^{-1}$ with a drastic drop-off in linear size beyond 3 Mpc. This indicates that GRGs are large, low surface brightness sources that show a decrease in radio power with the increase in size, as expected for luminosity evolution in active radio galaxies \citep{1997MNRAS.292..723K,1999Natur.399..330B}. Furthermore, this cut-off is possibly limited by instrumental sensitivity.

\item The occurrence of GRGs with respect to redshift also shows (Figure~\ref{barplot}) a decrease thereby indicating limitation in detection of distant GRGs with current surveys. It also highlights the need for high sensitivity,  low frequency radio surveys in the near future with telescopes like uGMRT, LOFAR and SKA.
\end{enumerate}

\vspace{-0.2in}
\section*{Acknowledgement}
We acknowledge the comments and suggestions of the anonymous referee which has helped us improve the quality of our paper. PD, JB and MP gratefully acknowledge generous support from the Indo-French Center for the Promotion of Advanced Research (Centre Franco-Indien pour la Promotion de la Recherche Avan\'{c}ee) under programme no. 5204-2. We thank the people who made NVSS, FIRST $\&$ SUMSS for the radio data and SDSS $\&$ 6dF for optical data. We thank IUCAA (especially Radio Physics Lab), Pune for providing all the facilities for carrying out the research work. We thank late Dr. Homi Bhabha for the inspiration. We thank Mr Tejas Kale for discussion related statistical tests. We gratefully acknowledge the use of Edward (Ned) Wright's online Cosmology Calculator. This research has made use of the NASA Extragalactic Database (NED) which is operated by the Jet Propulsion Laboratory, California Institute of Technology, under contract with the National Aeronautics and Space Administration. This research has also made use of the SIMBAD database, operated at CDS, Strasbourg, France. This publication makes use of data products from the Wide-field Infrared Survey Explorer, which is a joint project of the University of California, Los Angeles, and the Jet Propulsion Laboratory/California Institute of Technology, funded by the National Aeronautics and Space Administration.

%%%%%%%%%%%%%%%%%%%%%%%%%%%%%%%%%%%%%%%%%%%%%%%%%%


%\bibliographystyle{mnras}
\bibliographystyle{mnras} 
\bibliography{GRG_Dabhade}

%%%%%%%%%%%%%%%%% APPENDICES %%%%%%%%%%%%%%%%%%%%%

\appendix
\section{Some extra material}
%------------------------------------------------------------------------------------------------------------------------
\begin{figure*}
\vbox{
\hbox{
\hspace{0.5cm}
\subfloat[\textbf{GRG1 spectrum}\label{spec1}]{%
\includegraphics[height=2.5in,width=3.2in]{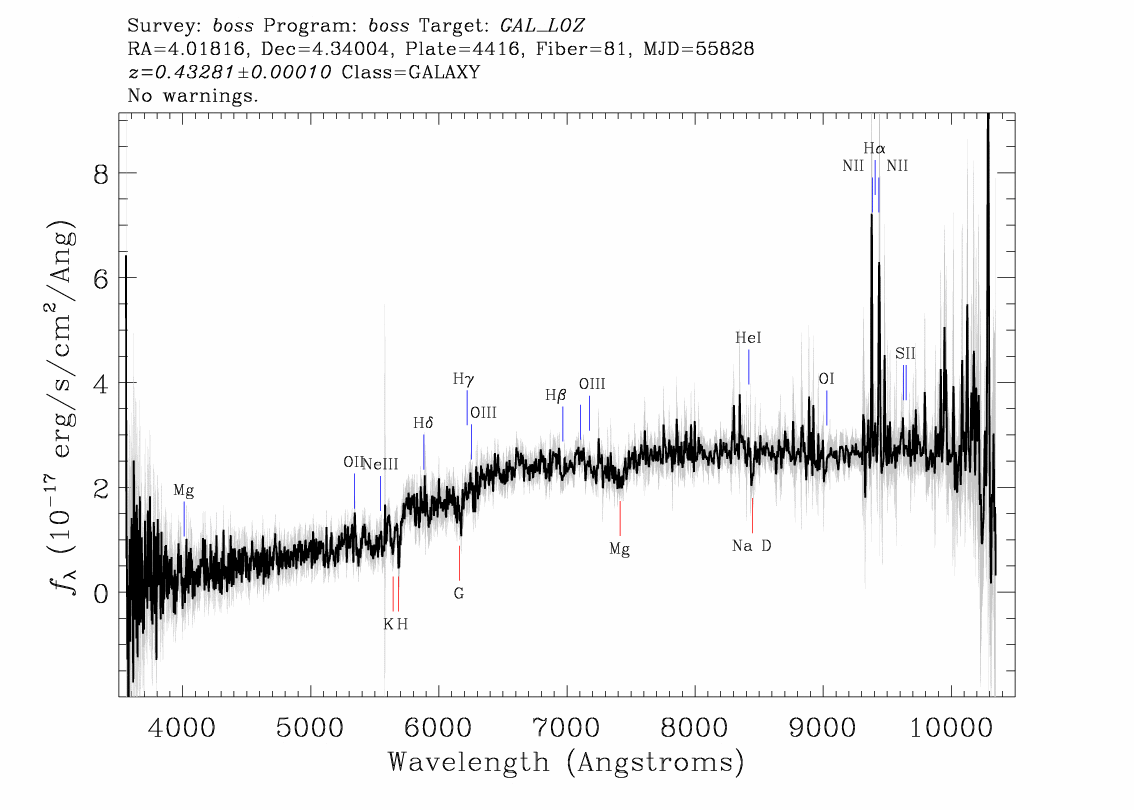}
}
\hspace{1.0cm}
\subfloat[\textbf{GRG2 spectrum}\label{spec2}]{%
\includegraphics[height=2.5in,width=3.2in]{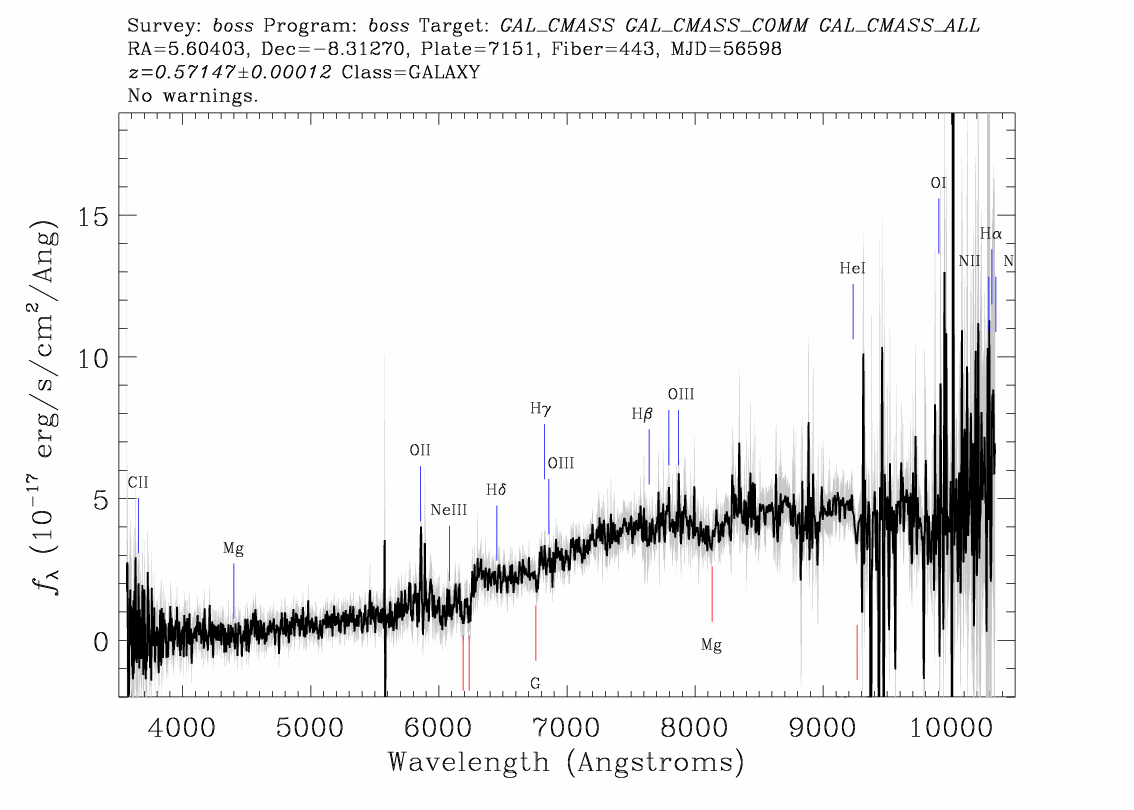}
}
}
\hbox{
\hspace{1.0cm}
\subfloat[\textbf{GRG7 spectrum}\label{spec7}]{%
\includegraphics[height=2.5in,width=3.2in]{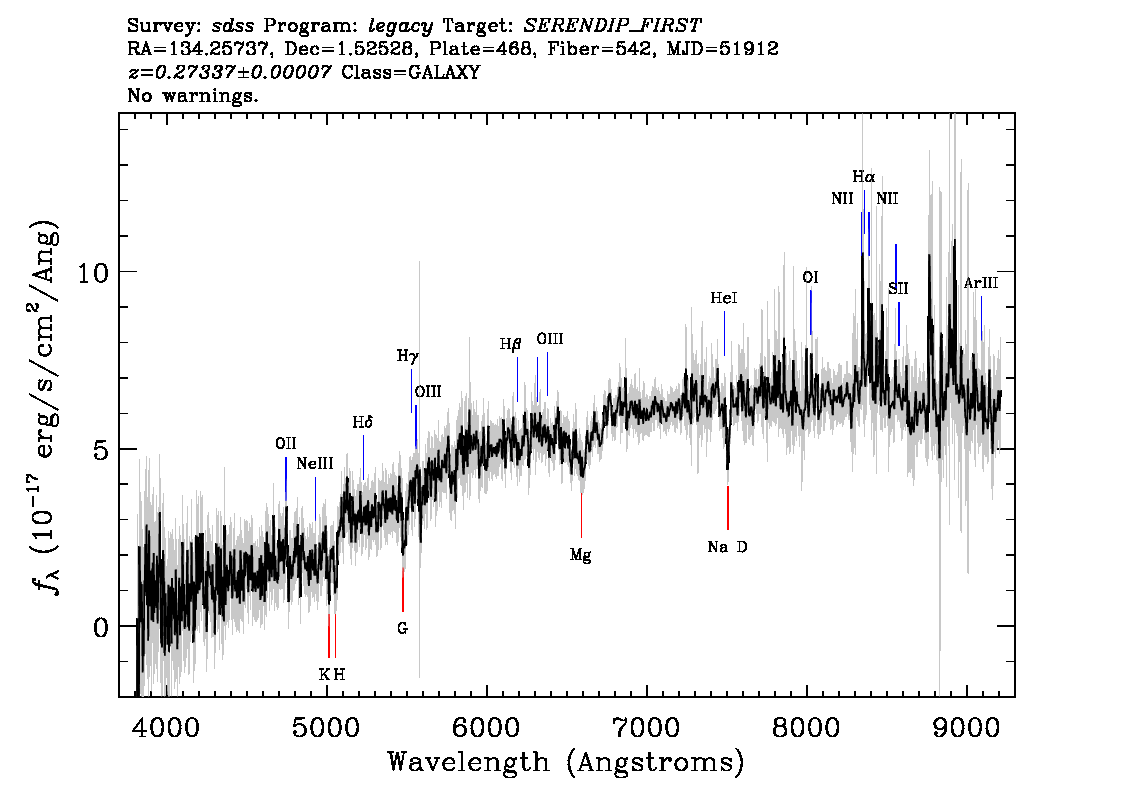}
}
\hspace{0.5cm}
\subfloat[\textbf{GRG10 spectrum}\label{spec10}]{%
\includegraphics[height=2.5in,width=3.2in]{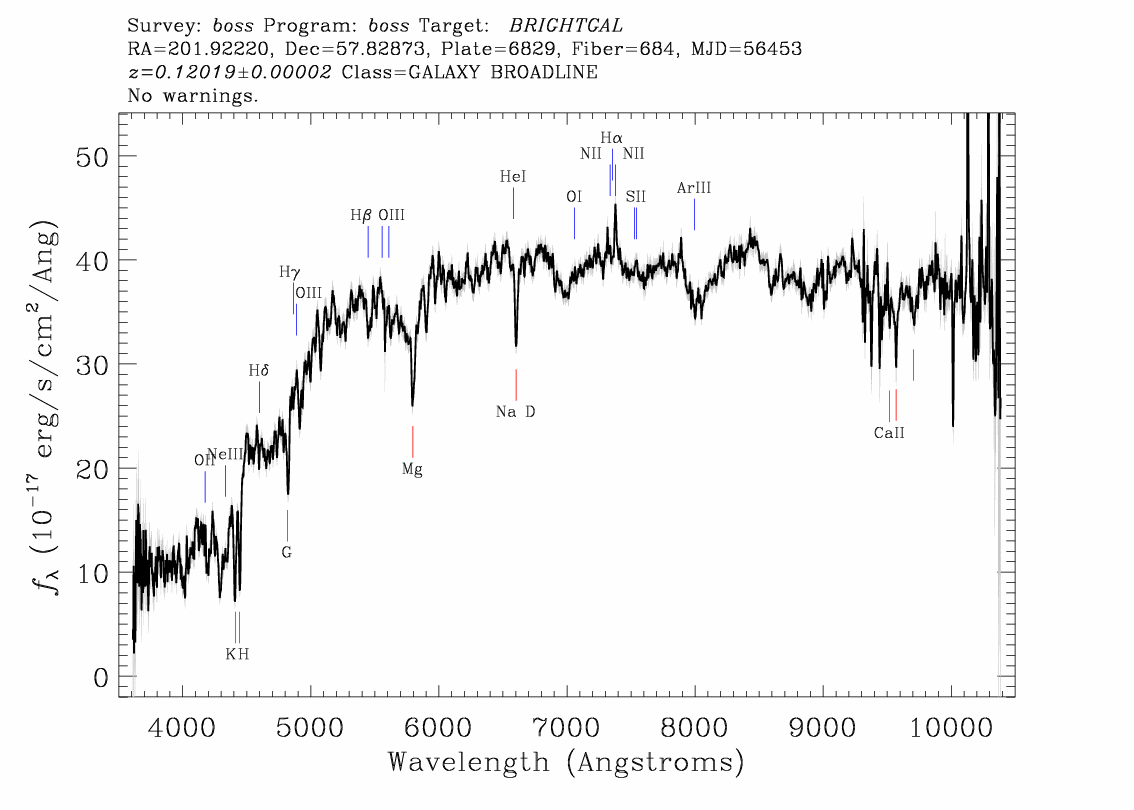}
}
}
\hbox{
\hspace{0.5cm}
\subfloat[\textbf{GRG19 spectrum}\label{spec19}]{%
\includegraphics[height=2.5in,width=3.2in]{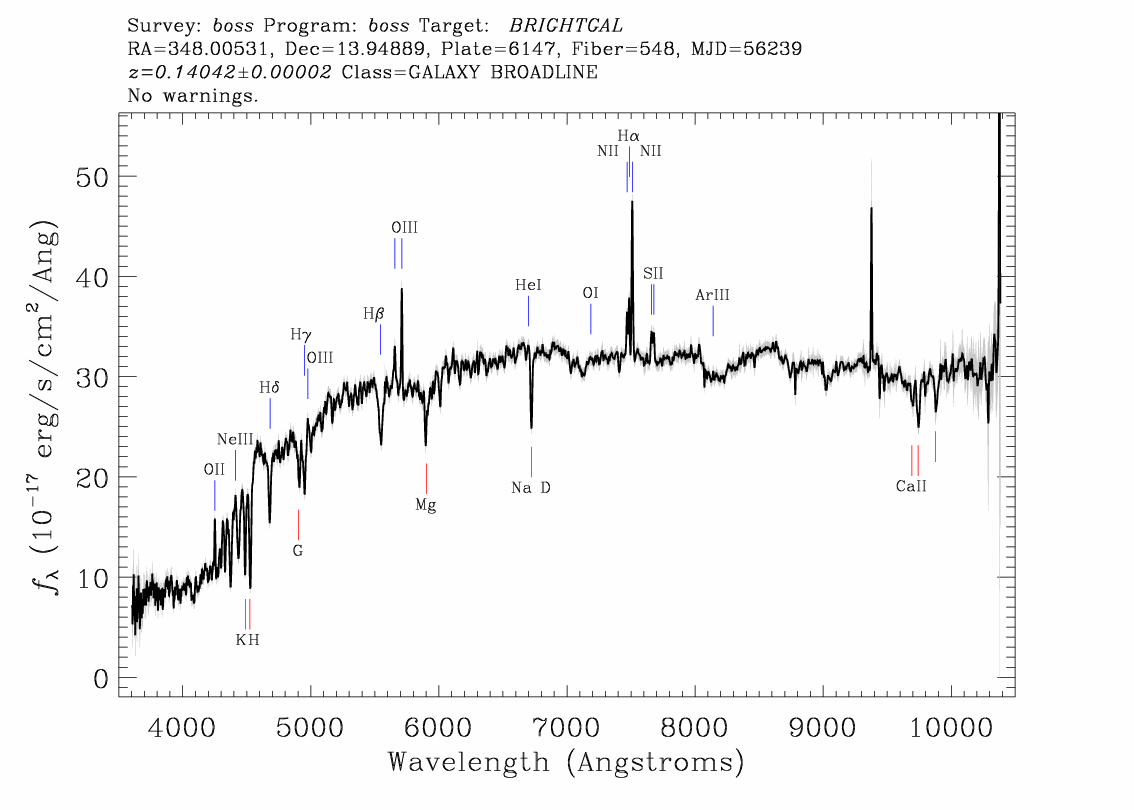}
}
\hspace{1.0cm}
\subfloat[\textbf{GRG21 spectrum}\label{spec21}]{%
\includegraphics[height=2.5in,width=3.2in]{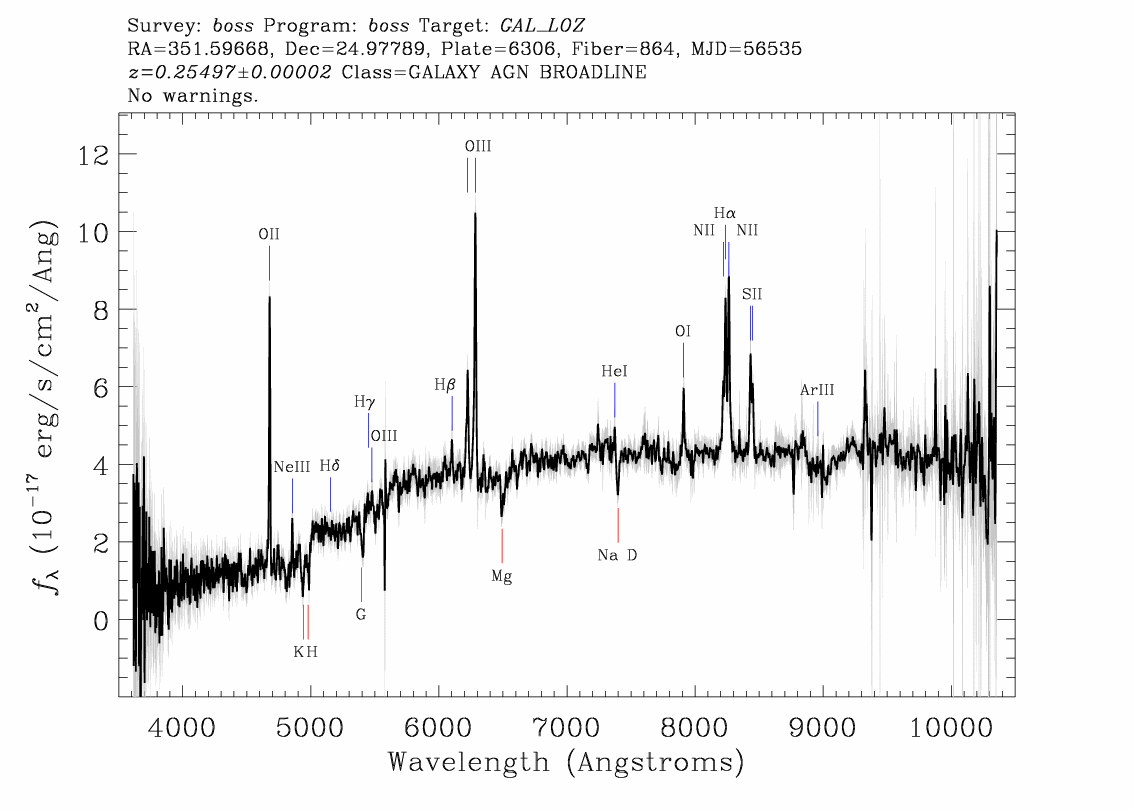}
}
}
\vspace{0.2in}
}
\caption{Spectra of hosts of GRG1, GRG2, GRG7, GRG10, GRG19 and GRG21 obtained from SDSS DR12 sky server.}
\label{spectrum}
\end{figure*}
%------------------------------------------------------------------------------------------------------------------------

%------------------------------------------------------------------------------------------------------------------------
\begin{figure}
\centering
\includegraphics[height=6.5in,width=7.5in,angle =90]{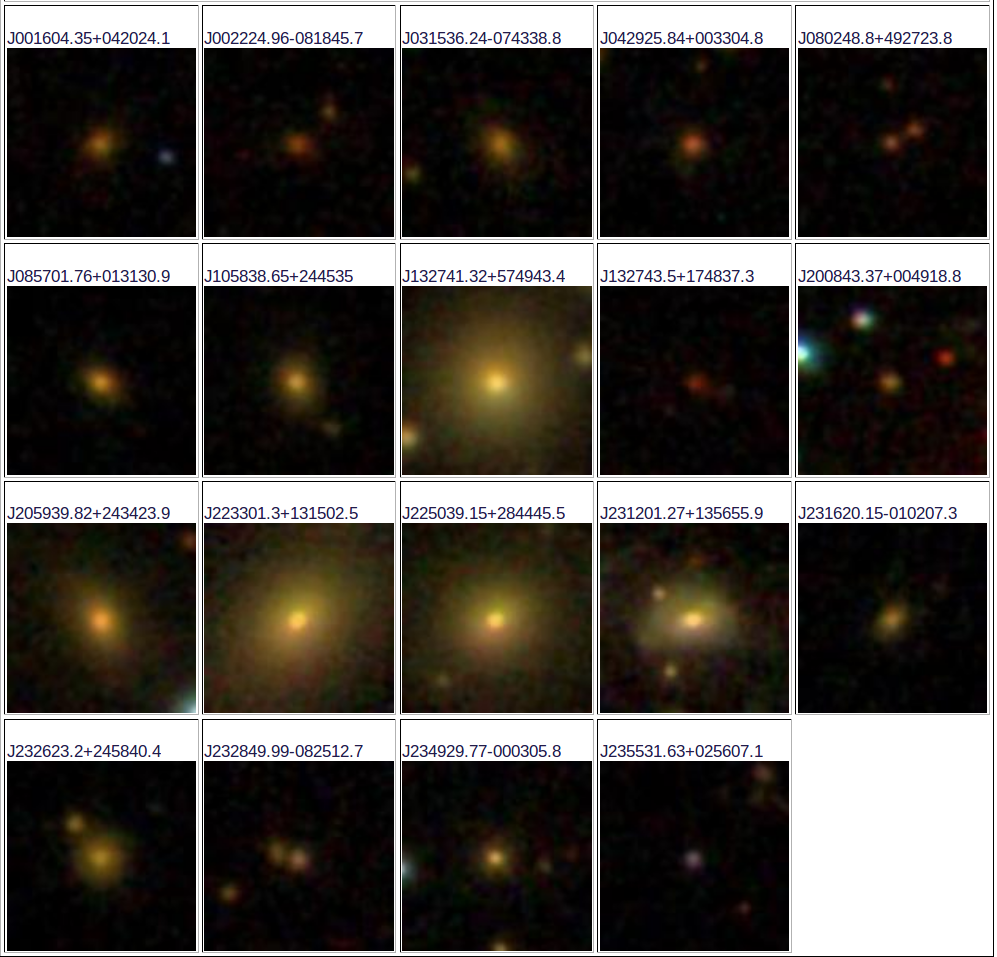}
\caption{Optical images of host galaxies of GRGs from SDSS.}
\label{optical}
\end{figure}
%------------------------------------------------------------------------------------------------------------------------

%%%%%%%%%%%%%%%%%%%%%%%%%%%%%%%%%%%%%%%%%%%%%%%%%%

% Don't change these lines
\bsp	% typesetting comment
\label{lastpage}

\end{document}